\documentclass[12pt]{article}
\usepackage{amsmath,amsfonts,amssymb}
\usepackage{hyperref}
\usepackage{color}
 
\def\bS{\mathbf{S}}
\allowdisplaybreaks
\usepackage{graphicx}
\newcommand{\beq}{\begin{equation}}
\newcommand{\eeq}{\end{equation}}
\newcommand{\beqa}{\begin{eqnarray}}
\newcommand{\eeqa}{\end{eqnarray}}
\newcommand{\beqar}{\begin{eqnarray*}}
\newcommand{\eeqar}{\end{eqnarray*}}

\newcommand{\nn}{\nonumber}

\newcommand{\z}{\zeta}

\newcommand{\eg}{{\it e.g.,}\ }
\newcommand{\ie}{{\it i.e.,}\ }
\newcommand{\labell}[1]{\label{#1}} 
\newcommand{\reef}[1]{(\ref{#1})}
\usepackage{amsmath,amsfonts,amssymb}
\def\nn{\nonumber}
\def\bS{\boldsymbol{S}}

\begin{document}
\begin{titlepage}
\begin{center}

\vskip 2 cm
{\LARGE \bf  T-duality invariant effective actions\\  \vskip 0.25 cm at orders $ \alpha', \alpha'^2$  
 }\\
\vskip 1.25 cm
 Hamid Razaghian\footnote{razaghian.hamid@gmail.com}  and  Mohammad R. Garousi\footnote{garousi@um.ac.ir}

\vskip 1 cm
{{\it Department of Physics, Faculty of Science, Ferdowsi University of Mashhad\\}{\it P.O. Box 1436, Mashhad, Iran}\\}
\vskip .1 cm
 \end{center}

\begin{abstract}
We use compatibility of  the $D$-dimensional effective actions   for diagonal metric and for dilaton  with the  T-duality when theory is compactified on a circle,   to find  the the $D$-dimensional couplings of curvatures and dilaton as well as the higher derivative corrections to the $(D-1)$-dimensional Buscher rules  at orders $ \alpha' $ and $\alpha'^2$.  We observe that the T-duality constraint on the effective actions   fixes   the covariant effective actions at each order of $\alpha'$ up to   field redefinitions and up to an overall factor. Inspired by these results, we   speculate that the $D$-dimensional effective actions at any order of $\alpha'$ must be consistent with the standard Buscher rules provided that one uses covariant field redefinitions in the corresponding reduced $(D-1)$-dimensional effective actions. This constraint may be used to find effective actions at all higher orders of  $\alpha'$.
\end{abstract}
\end{titlepage}

\section{Introduction}

String theory is a field theory with  a finite number of massless fields and a  tower of infinite number of  massive fields. An efficient way to study different phenomena in this theory is to use an effective action in which  effects of the massive fields appear in higher derivatives of the massless fields. There are verity of methods for finding such higher derivative couplings. S-matrix element approach \cite{Scherk:1974mc,Yoneya:1974jg}, sigma-model approach \cite{Callan:1985ia,Fradkin:1984pq,Fradkin:1985fq},  supersymmetry approach \cite{Gates:1986dm,Gates:1985wh,Bergshoeff:1986wc}, Double Field Theory (DFT) approach  \cite{Siegel:1993xq,Hull:2009mi,Hohm:2010jy},  and duality approach \cite{Ferrara:1989bc,Font:1990gx,Garousi:2017fbe}. In the duality approach, the  consistency of the effective actions with duality transformations are imposed to find the higher derivative couplings \cite{Garousi:2017fbe}. In this paper, we are interested in constraining the couplings to be consistent with T-duality transformations.

The T-duality in string theory is realized by studying the spectrum of the closed string on a tours. The spectrum is invariant under the transformation in which the Kaluza-Kelin modes and the winding modes are interchanged, and at the same time the set of scalar fields parametrizing the tours  transforms to another set of scalar fields parametrizing  the dual tours.  The transformations on the scalar fields have been extended to curved spacetime with background fields by Buscher \cite{Buscher:1987sk,Buscher:1987qj}. It has been observed that the effective actions at the leading order of $\alpha'$ are invariant  under the Buscher rules \cite{Sen:1991zi}. When this idea is going to be implemented  on the higher derivative corrections to the effective actions, however, one runs into the problem that the Buscher rules should receive higher derivative corrections as well \cite{Kaloper:1997ux}. 

There are covariant and non-covariant  approaches for applying the T-duality on the effective actions at the higher order of $\alpha'$ \cite{Garousi:2017fbe}. In the non-covariant approach,      non-covariant effective actions are  constrained to be invariant under the standard Buscher rules with no higher derivative correction \cite{Godazgar:2013bja,Garousi:2015qgr}. Since the effective actions   are not covariant under the general coordinate transformations, some  non-covariant field redefinitions are required to convert the effective actions to the covariant forms. The field redefinitions may cause   the Buscher rules to receive higher covariant derivative corrections as well.  Alternatively, in the   covariant approach in which we are interested in this paper,   one considers     covariant effective actions and then requires them to be invariant under the appropriate T-duality transformations which are the Buscher rules plus their corresponding higher covariant derivative corrections.

In the bosonic   and in the heterotic string theories, the higher derivative corrections to the effective actions begin at order $\alpha'$, whereas, in type II superstring theory, the   higher derivative corrections begin at order $\alpha'^3$. As a result, the   corrections to the Buscher rules in the covariant approach begin at order $\alpha'$ in the bosonic and heterotic theories, and begin at order $\alpha'^3$ in the superstring theory. Since there is  no higher derivative correction at order $\alpha'^2$ to the Buscher rules in the superstring theory, a covariant   effective action  for O$_p$-plane at order $\alpha'^2$ has been found in \cite{Robbins:2014ara,Garousi:2014oya} by requiring it to be consistent with the  standard Buscher rule. 
In this paper, we are interested in $\alpha',\, \alpha'^2$ corrections to the bulk effective actions of the bosonic and heterotic theories. Hence, in the covariant approach, we have to work with the Buscher rules plus their $\alpha'$ and $ \alpha'^2$ corrections. 

Using   effective action of the bosonic and heterotic string theories at order $\alpha'$ in an specific field variables, Kaloper and Meissner have found corrections at order $\alpha'$ to the Buscher rules that make the effective action to be  consistent with T-duality  \cite{Kaloper:1997ux}. 
In this paper, we are going to answer the following  question:  If one considers an effective action with all possible independent covariant couplings at  each order of $\alpha'$ with unknown coefficients and considers the  Buscher rules plus all possible  higher covariant derivative corrections at the same order of $\alpha'$, then, is the requirement that the action to be consistent with the T-duality transformation is    strong enough to fix   the coefficients in the  effective action and in the T-duality transformations?   At orders $\alpha'$ and $\alpha'^2$, and  for the simple case  that there is no B-field and  the metric is diagonal that we have done the calculations, we have found that the T-duality constraint fixes the action up  to  field redefinitions and up to an overall factor. It also   fixes  many unknown coefficients in  the T-duality transformations. There are, however, some residual   T-duality parameters that may be fixed by the calculations in the presence of B-field or by the calculations at the higher orders of $\alpha'$.

The T-duality transformations at the leading order of $\alpha'$ are given by the Buscher rules. They form a $ \mathbb{Z}_2 $ group. One may impose this condition on the T-duality transformation at the higher order of $\alpha'$ to constrain  the higher derivative corrections to the Buscher rules. As we will see, this condition  excludes many covariant derivative terms in the T-duality transformations and interrelate coefficients of many other terms.  Furthermore, the criterion  that  the effective actions are invariant under the T-duality transformations relate the non-zero coefficients in the T-duality transformations  to the coefficients of the effective actions.    Using these conditions, one may  find both the effective actions and the T-duality transformations at any order of  $\alpha'$.  

In general, apart from  coefficients of the Riemann curvature couplings which are independent of field redefinitions, the coefficients of all other couplings in the effective actions are a priori ambiguous. The ambiguous coefficients transform under the field redefinitions $ G_{\mu\nu}\to G_{\mu\nu}+\alpha' \delta G^{(1)}_{\mu\nu}+\cdots$ and $ \Phi\to \Phi+\alpha'\delta\Phi^{(1)}+\cdots $. So there are a large number of effective actions, which all are physically identical, \ie all correspond to the same S-matrix elements. At order $\alpha'$, it has been shown in  \cite{Metsaev:1987zx}  that    there are  8 ambiguous coefficients and they  satisfy one relation which is invariant under the field redefinitions. So one can set all the ambiguous coefficients to arbitrary numbers  except one of them. The S-matrix calculation can fix the remaining coefficient. At order $\alpha'^2$, there are 42 ambiguous coefficients. They satisfy 5 relations which are invariant under the field redefinitions \cite{Bento1989,Bento1990,Bento:1990jc}.  So one can fix all ambiguous coefficients to arbitrary numbers except 5 of them. The unfixed coefficients can be found from the corresponding S-matrix elements. Similarly for couplings at higher order of $\alpha'$. The combinations of the ambiguous coefficients that remain invariant under the field redefinitions may be     functions of the coefficients of the Riemann curvature terms which are also invariant under the field redefinitions.  Theses functions may be found from the S-matrix calculations. In this paper, among other things, we are going to find these functions from the T-duality constraints on the  effective actions.  We will find that these functions are all in fact zero. That is, there is no combination of the coefficients of the T-duality invariant theory which remains invariant under field redefinitions. This indicates that the T-duality transformations do not relate the coefficients of the Riemann curvature couplings to   all other couplings.   For example, as we will see, the T-duality fixes the effective action at order $\alpha'$ to be 
\begin{align}
& S_1=\frac{-2}{\kappa}\alpha'\int d^{d+1}x e^{-2\Phi}\sqrt{-G}\Big( b_1 R_{\alpha \beta \gamma \delta} R^{\alpha \beta \gamma \delta}+ b_2 R_{\alpha \beta} R^{\alpha \beta} + b_3 R^2  \nonumber\\
& +   b_4 R_{\alpha \beta} \nabla^{\alpha}\Phi \nabla^{\beta}\Phi+  b_5 R \nabla_{\alpha}\Phi \nabla^{\alpha}\Phi +b_6 R \nabla_{\alpha}\nabla^{\alpha}\Phi +   b_7 \nabla_{\alpha}\nabla^{\alpha}\Phi \nabla_{\beta}\nabla^{\beta}\Phi \nn \\ 
&+ b_8 \nabla_{\alpha}\Phi \nabla^{\alpha}\Phi \nabla_{\beta}\nabla^{\beta}\Phi   -2 (8 b_{3} - 2 b_{5} - 4 b_{6} + 2 b_{7} + b_{8}) \nabla_{\alpha}\Phi \nabla^{\alpha}\Phi \nabla_{\beta}\Phi \nabla^{\beta}\Phi\Big)\labell{finalS10}
\end{align}
where the coefficients $b_1,b_2,\cdots, b_8$ are all arbitrary.   The corresponding T-duality corrections to the Buscher rules  appear in \reef{Trans1}. 
Note that in the effective action \reef{finalS10},  the coefficient of the   Riemann curvature coupling does not relate to any other coupling. Moreover, there are 
  no ambiguous coefficients in the action any more, \ie there is no combination of the  coefficients $b_2,\cdots, b_8$ which is invariant nuder the field redefinitions. Different choices for these  coefficients give the effective action in different schemes.  For example, the effective action   in the ``Gauss-Bonnet" scheme   is
\begin{align}
S_1= \frac{-2b_1}{\kappa^2}\alpha'\int  e^{-2\Phi}\sqrt{-G}\Bigl( R_{\alpha \beta \gamma \delta}R^{\alpha \beta \gamma \delta}-4R_{\alpha\beta}R^{\alpha\beta}+R^2-16 \nabla_{\alpha}\Phi \nabla^{\alpha}\Phi \nabla_{\beta}\Phi \nabla^{\beta}\Phi  \Bigr)\labell{GB}
\end{align}
  This action   is consistent with the S-matrix calculation \cite{Metsaev:1986yb}.

One may consider the higher covariant derivative corrections \reef{Trans1} to the Buscher rules    as field redefinitions in the reduced $(D-1)$-dimensional effective action.  If one is not interested in the explicit form of the field redefinitions at order $\alpha'$, one may substitute the equations of motion at order $\alpha'^0$ to the effective action at order $\alpha'$ (see \eg \cite{Meissner:1996sa}). We will find the T-duality invariant effective action \reef{finalS10} in this way as well. However, at higher orders of $\alpha'$, in general,  making the field redefinitions is not equivalent to  all possible substitutions of the lower order equations of motion in the  effective action at order $\alpha'^n$. 

The outline of the paper is as follows: In section 2, we consider the most general T-duality transformations at order $\alpha',\alpha'^2$, and use the fact that the T-duality transformations must form a $ \mathbb{Z}_2 $ group, to exclude some of the terms in the transformations and to find some relations between the non-zero terms.    In section 3, we use the compatibility of a priori unknown effective actions with the T-duality transformations   to fix both the effective actions and their corresponding  T-duality transformations. In subsection 3.1, we show that the effective action at the leading order of $\alpha'$ is fixed up to an overall factor. In subsection 3.2, we show that the effective action at order $\alpha'$ is fixed up to an overall factor and up to field redefinitions. The corresponding T-duality transformations at order $\alpha'$ are also found in this section.   We show that for one particular field variables,   the effective action and its corresponding T-duality transformations   are those    appear in the literature. In subsection 3.3, we   repeat   the calculations to the order $\alpha'^2$ and find the effective action  and its corresponding   T-duality transformations. The effective action is again fixed up to an overall factor and up to field redefinitions. In two  particular field variables, we show that the effective actions are those   appear in the literature which have been  found from the S-matrix and the sigma-model calculations.  

\section{$ \mathbb{Z}_2 $-constraint on T-duality transformations}

In this section, we are going to consider the T-duality transformations which include all higher derivative terms at each order of $\alpha'$ with unknown coefficients and constrain the coefficients such that they form  a $ \mathbb{Z}_2 $ group.  To simplify the calculations,   we are going to consider the   case that  the theory is compactified on a circle with the killing coordinate $y$ and radius $\rho$, \ie $D=d+1$ where $D$ is the dimension of spacetime. In this case, the Buscher rules which are the T-duality transformations at the leading order of $\alpha'$, are \cite{ Buscher:1987sk,Buscher:1987qj}  
\beqa
e^{2{ \Phi}}\stackrel{T^{(0)}}{\longrightarrow}\frac{e^{2\Phi}}{G_{yy}}&;& 
G_{yy}\stackrel{T^{(0)}}{\longrightarrow}\frac{1}{G_{yy}}\nonumber\\
G_{a y}\stackrel{T^{(0)}}{\longrightarrow}\frac{B_{a y}}{G_{yy}}&;&
G_{ab}\stackrel{T^{(0)}}{\longrightarrow}G_{ab}-\frac{G_{a y}G_{b y}-B_{a y}B_{b y}}{G_{yy}}\nonumber\\
B_{a y}\stackrel{T^{(0)}}{\longrightarrow}\frac{G_{a y}}{G_{yy}}&;&
B_{ab}\stackrel{T^{(0)}}{\longrightarrow}B_{ab}-\frac{B_{a y}G_{b y}-G_{a y}B_{b y}}{G_{yy}}\labell{nonlinear}
\eeqa
where $a,b$ denote any   direction other than $y$. In above transformation, the metric is  in the string frame. It is easy to verify that the above transformations form a $ \mathbb{Z}_2 $ group, \ie
\beqa
e^{2\Phi}\stackrel{T^{(0)}}{\longrightarrow}\frac{e^{2\Phi}}{G_{yy}}\stackrel{T^{(0)}}{\longrightarrow}e^{2\Phi}&;& 
G_{yy}\stackrel{T^{(0)}}{\longrightarrow}\frac{1}{G_{yy}}\stackrel{T^{(0)}}{\longrightarrow}G_{yy}\nonumber\\
G_{a y}\stackrel{T^{(0)}}{\longrightarrow}\frac{B_{a y}}{G_{yy}}\stackrel{T^{(0)}}{\longrightarrow}G_{a y}&;&
G_{ab}\stackrel{T^{(0)}}{\longrightarrow}G_{ab}-\frac{G_{a y}G_{b y}-B_{a y}B_{b y}}{G_{yy}}\stackrel{T^{(0)}}{\longrightarrow}G_{ab}\nonumber\\
B_{a y}\stackrel{T^{(0)}}{\longrightarrow}\frac{G_{a y}}{G_{yy}}\stackrel{T^{(0)}}{\longrightarrow}B_{a y}&;&
B_{ab}\stackrel{T^{(0)}}{\longrightarrow}B_{ab}-\frac{B_{a y}G_{b y}-G_{a y}B_{b y}}{G_{yy}}\stackrel{T^{(0)}}{\longrightarrow}B_{ab}\nonumber
\eeqa
This property must be carried out by all the  higher derivative corrections  to the Buscher rules.

The Buscher rules are the T-duality transformations corresponding to the effective action at the leading order of $\alpha'$. The  T-duality transformations corresponding to the effective action at all orders of  $\alpha'$ may have   the following  $\alpha'$-expansion:
\beqa
T&=&\sum_{n=0}^{\infty}(\alpha')^nT^{(n)}\,,\labell{T}
\eeqa
where $T^{(0)}$ is the above Buscher rules, $T^{(1)}$ is correction to the Buscher rules at order $\alpha'$, and so on.   The fact that the T-duality transformation \reef{T} must  form a $ \mathbb{Z}_2 $ group, produces schematically the following constraints:
\beqa
T^{(0)}T^{(0)}&=&1,\nonumber\\
T^{(0)}T^{(1)}&=&0,\nonumber\\
T^{(0)}T^{(2)}+T^{(1)}T^{(1)}&=&0,\nonumber\\
T^{(0)}T^{(3)}+T^{(1)}T^{(2)}&=&0,\nonumber\\
T^{(0)}T^{(4)}+T^{(1)}T^{(3)}+T^{(2)}T^{(2)}&=&0, \labell{TT}\\
\vdots&&\nonumber
\eeqa
The first relation is the $ \mathbb{Z}_2 $-constraint at order $\alpha'^0$, the second one is  the $ \mathbb{Z}_2 $-constraint at order $\alpha'$, the third one  is  the $ \mathbb{Z}_2 $-constraint at order $\alpha'^2$, and so on.   One can write the correction $T^{(n)}$ in terms of all possible d-dimensional tensors at order $\alpha'^n$ with unknown coefficients. Then the above relations   may be used to constrain the unknown coefficients in  $T^{(1)},T^{(2)},\cdots$.

For the simple case that the metric is diagonal and B-field is zero, the Buscher rules become
\beqa
e^{2\Phi}\stackrel{T^{(0)}}{\longrightarrow}\frac{e^{2\Phi }}{G_{yy}}&;&G_{yy}\stackrel{T^{(0)}}{\longrightarrow}\frac{1}{G_{yy}} 
\eeqa
and the $d$-dimensional metric $G_{ab}\equiv g_{ab}$  is invariant. Parametrizing the $d$-dimensional scalar $G_{yy}$ as $G_{yy}=e^{2\sigma}$, then the Buscher rules simplify to
\beqa
\sigma &\stackrel{T^{(0)}}{\longrightarrow}&-\sigma\nn\\
P &\stackrel{T^{(0)}}{\longrightarrow}&P\nn\\
g_{ab} &\stackrel{T^{(0)}}{\longrightarrow}&g_{ab}\labell{sigma}
\eeqa
 where $P$ is  the $d$-dimensional dilaton, \ie  $P=\Phi-\sigma/2$.

The corrections  at order $\alpha'$ to the Buscher rules in general   are:
\beqa
\sigma &\stackrel{T }{\longrightarrow}& -\sigma+\alpha'\delta\sigma^{1}+O(\alpha'^2)\nonumber\\
P &\stackrel{T }{\longrightarrow}& P+\alpha'\delta P^{1}+O(\alpha'^2)\labell{eq.2.correc}\\
g_{ab}&\stackrel{T }{\longrightarrow} &g_{ab}+\alpha'\delta g^{1}_{ab}+O(\alpha'^2)\nonumber
\eeqa
where $ \delta\sigma^{1}, \delta P^{1}$ and $ \delta g^{1}_{ab} $ are functions of $ \sigma,P , g_{ab} $  which have two $d$-dimensional covariant derivatives. To impose the constraint \reef{TT}, it is convenient to  separate $\delta\sigma,\, \delta P$ and $ \delta g_{ab}$ to  $  \sigma$-odd and  $  \sigma$-even parts. We then define the class A in which $\delta\sigma$ has even number of $\sigma$ and $\delta P,\delta g_{ab}$ have odd number of $\sigma$. All other terms are defind to be in class B, \ie
\beqa
\delta\sigma &=&\delta \sigma _A +\delta \sigma _B\nonumber\\
\delta P &=& \delta P _A +\delta P _B\nonumber\nonumber\\
\delta g_{ab} &=&\delta g_{Aab} +\delta g_{Bab}\labell{eq.2.2.201}
\eeqa
The corrections to the Buscher rule at order $\alpha'$ in class A are: 
\beqa
\delta\sigma^{1}_A(\sigma,P,g)&=&A_1 \tilde{R} + A_2 \tilde{\nabla}_{a}\tilde{\nabla}^{a}P  + A_3 \tilde{\nabla}_{a}P \tilde{\nabla}^{a}P + A_4 \tilde{\nabla}_{a}\sigma \tilde{\nabla}^{a}\sigma \nonumber\\
\delta P^{1}_A(\sigma,P,g)&=&A_5 \tilde{\nabla}_{a}\tilde{\nabla}^{a}\sigma + A_6 \tilde{\nabla}_{a}\sigma \tilde{\nabla}^{a}P  \nonumber\\
\delta g_{Aab}^{1}(\sigma,P,g)&=& A_7 (\tfrac{1}{2} \tilde{\nabla}_{a}\sigma \tilde{\nabla}_{b}P+ \tfrac{1}{2} \tilde{\nabla}_{a}P \tilde{\nabla}_{b}\sigma)    + A_8 \tilde{\nabla}_{b}\tilde{\nabla}_{a}\sigma \nonumber \\
&&+g_{ab}\Big( A_9\tilde{\nabla}_{c}\tilde{\nabla}^{c}\sigma  
 + A_{10}  \tilde{\nabla}_{c}\sigma \tilde{\nabla}^{c}P \Big) \labell{eq.2.2.20}
\eeqa
where $A_1,\ldots A_{10}$  are some unknown coefficients. The tilde sign over the covariant derivatives means the metric in them is the $d$-dimensional metric $g_{ab}$. One should exclude  the transformations that are correspond to the $d$-dimensional coordinate transformations. Under infinitesimal coordinate transformation $x^a\rightarrow x^a-\z^a(x)$, the metric $g_{ab}$, the  dilaton $P$ and $\sigma$ transform as $\delta g_{ab}=\tilde{\nabla}_a\z_b+\tilde{\nabla}_b\z_a$, $\delta P=\z^a\tilde{\nabla}_a P$ and $\delta \sigma=\z^a\tilde{\nabla}_a\sigma$. If one chooses the infinitesimal parameter as $\z^a=(A_8/2)\tilde{\nabla}^{a}\sigma$, then the corresponding coordinate transformations are $\delta g_{ab}=A_8\tilde{\nabla}_a\tilde{\nabla}_b\sigma$, $\delta P=(A_8/2)\tilde{\nabla}^{a}\sigma\tilde{\nabla}_a P$ and $\delta \sigma=(A_8/2)\tilde{\nabla}^{a}\sigma\tilde{\nabla}_a \sigma$. Therefore, one should exclude the term $A_8\tilde{\nabla}_a\tilde{\nabla}_b\sigma$ in $\delta g_{Aab}^{1}(\sigma,P,g)$. Since we are working in covariant approach, the presence of such term should not however affect our calculations, so we keep the term with coefficient $A_8$ and set it to zero at the end of our calculations.
 
Using the $ \mathbb{Z}_2 $-constraints at order $\alpha'$, one finds $\delta\sigma^{1}_B,\delta P^{1}_B,\delta g_{Bab}^{1}$ are all zero. That is, the $ \mathbb{Z}_2 $-constraints are
\begin{align}
&\sigma\xrightarrow{T}-\sigma+\alpha'(\delta\sigma^1_A+\delta\sigma^1_B)+O(\alpha'^2)\xrightarrow{T} \sigma-2\alpha'\delta\sigma^1_B+O(\alpha'^2)=\sigma\nn\\
&P\xrightarrow{T}P+\alpha'(\delta P^1_A+\delta P^1_B)+O(\alpha'^2)\xrightarrow{T}P+2\alpha'\delta P^1_B+O(\alpha'^2)=P\nn\\
&g_{ab}\xrightarrow{T}g_{ab}+\alpha'(\delta g_{Aab}^1+\delta g_{Bab}^1)+O(\alpha'^2)\xrightarrow{T}g_{ab}+2\alpha'\delta g_{Bab}^1+O(\alpha'^2)=g_{ab}\nn\, .
\end{align}
where we have used the fact that 
\beqa
\delta\sigma_A(-\sigma)=\delta\sigma_A(\sigma)&;&\delta P_A(-\sigma)=-\delta P_A(\sigma)\,\,\,;\,\,\,\delta g_{Aab}(-\sigma)=-\delta g_{Aab}(\sigma)\nn\\
\delta\sigma_B(-\sigma)=-\delta\sigma_B(\sigma)&;&\delta P_B(-\sigma)=\delta P_B(\sigma)\,\,\,;\,\,\,\delta g_{Bab}(-\sigma)= \delta g_{Bab}(\sigma)\labell{con1}
\eeqa
So the only non-zero terms at order $\alpha'$ are those in \reef{eq.2.2.20}.

We extend the above calculations to one higher order of  $ \alpha' $, \ie
\beqa
\sigma &\stackrel{T }{\longrightarrow} & -\sigma+\alpha'\delta\sigma^{1}_A+\frac{1}{2}\alpha'^2(\delta\sigma^{2}_A+\delta\sigma^{2}_B)+O(\alpha'^3)\nonumber\\
P &\stackrel{T }{\longrightarrow} & P+\alpha'\delta P^{1}_A+\frac{1}{2}\alpha'^2(\delta P^{2}_A+\delta P^{2}_B)+O(\alpha'^3)\labell{eq.2.correc.2}\\
g_{ab} &\stackrel{T }{\longrightarrow} & g_{ab}+\alpha'\delta g^{1}_{Aab}+\frac{1}{2}\alpha'^2(\delta g^{2}_{Aab}+\delta g^{2}_{Bab})+O(\alpha'^3)\nonumber
\eeqa
where $ \delta\sigma^{2}, \delta P^{2}$ and $ \delta g^{2}_{ab} $ are functions of $ \sigma,P , g_{ab} $  which have four $d$-dimensional covariant derivatives.
Using the symmetries \reef{eq.2.2.20}, the $ \mathbb{Z}_2 $-constraint now leaves $\delta\sigma_A^2,\delta p_A^2, \delta g_{Aab}^2$ arbitrary and fixes $\delta\sigma_B^2,\delta p_B^2, \delta g_{Bab}^2$ as
\beqa
\alpha'^2\delta\sigma_B^2&=&\alpha'\delta\sigma_A^1(-\sigma+\alpha'\delta\sigma_A^1,P+\alpha'\delta P_A^1,g+\alpha'\delta g_A^1)-\alpha'\delta\sigma_A^1(\sigma,P,g)\nn\\
\alpha'^2\delta P_B^2&=&-\alpha'\delta P_A^1(-\sigma+\alpha'\delta\sigma_A^1,P+\alpha'\delta P_A^1,g+\alpha'\delta g_A^1)-\alpha'\delta P_A^1(\sigma,P,g)\nn\\
\alpha'^2\delta g_{Bab}^2&=&-\alpha'\delta g_{Aab}^1(-\sigma+\alpha'\delta\sigma_A^1,P+\alpha'\delta P_A^1,g+\alpha'\delta g_A^1)-\alpha'\delta\ g_{Aab}^1(\sigma,P,g)
\eeqa
Using \reef{eq.2.2.20}, one can write $\delta\sigma_B^2,\delta p_B^2, \delta g_{Bab}^2$ in terms of product of two $A_1,\cdots A_{10}$. 
 The arbitrary corrections $\delta\sigma_A^2,\delta p_A^2, \delta g_{Aab}^2$ are: 
\begin{align}
\delta \sigma^2_A=&
A_{11} \tilde{R}_{ab} \tilde{R}^{ab} + A_{12} \tilde{R}^2 + A_{13} \tilde{R}_{abcd} \tilde{R}^{abcd} + A_{14} \tilde{R} \tilde{\nabla}_{a}\tilde{\nabla}^{a}P + A_{15} \tilde{\nabla}_{a}\tilde{\nabla}^{a}\tilde{R} + A_{16} \tilde{R} \tilde{\nabla}_{a}P \tilde{\nabla}^{a}P \nonumber \\ 
& + A_{17} \tilde{\nabla}_{a}\tilde{R} \tilde{\nabla}^{a}P + A_{18} \tilde{R} \tilde{\nabla}_{a}\sigma \tilde{\nabla}^{a}\sigma + A_{19} \tilde{R}^{ab} \tilde{\nabla}_{b}\tilde{\nabla}_{a}P + A_{20} \tilde{\nabla}_{a}\tilde{\nabla}^{a}P \tilde{\nabla}_{b}\tilde{\nabla}^{b}P \nonumber \\ 
& + A_{21} \tilde{\nabla}_{a}P \tilde{\nabla}^{a}P \tilde{\nabla}_{b}\tilde{\nabla}^{b}P + A_{22} \tilde{\nabla}_{a}\sigma \tilde{\nabla}^{a}\sigma \tilde{\nabla}_{b}\tilde{\nabla}^{b}P + A_{23} \tilde{\nabla}_{a}\tilde{\nabla}^{a}\sigma \tilde{\nabla}_{b}\tilde{\nabla}^{b}\sigma \nonumber \\ 
& + A_{24} \tilde{\nabla}_{a}\sigma \tilde{\nabla}^{a}P \tilde{\nabla}_{b}\tilde{\nabla}^{b}\sigma + A_{25} \tilde{\nabla}^{a}P \tilde{\nabla}_{b}\tilde{\nabla}^{b}\tilde{\nabla}_{a}P + A_{26} \tilde{\nabla}^{a}\sigma \tilde{\nabla}_{b}\tilde{\nabla}^{b}\tilde{\nabla}_{a}\sigma \nonumber \\ 
& + A_{27} \tilde{\nabla}_{b}\tilde{\nabla}^{b}\tilde{\nabla}_{a}\tilde{\nabla}^{a}P + A_{28} \tilde{R}_{ab} \tilde{\nabla}^{a}P \tilde{\nabla}^{b}P + A_{29} \tilde{\nabla}_{a}P \tilde{\nabla}^{a}P \tilde{\nabla}_{b}P \tilde{\nabla}^{b}P \nonumber \\ 
& + A_{30} \tilde{\nabla}_{a}\sigma \tilde{\nabla}^{a}P \tilde{\nabla}_{b}\sigma \tilde{\nabla}^{b}P + A_{31} \tilde{\nabla}^{a}P \tilde{\nabla}_{b}\tilde{\nabla}_{a}P \tilde{\nabla}^{b}P + A_{32} \tilde{R}_{ab} \tilde{\nabla}^{a}\sigma \tilde{\nabla}^{b}\sigma \nonumber \\ 
& + A_{33} \tilde{\nabla}_{a}P \tilde{\nabla}^{a}P \tilde{\nabla}_{b}\sigma \tilde{\nabla}^{b}\sigma + A_{34} \tilde{\nabla}_{a}\sigma \tilde{\nabla}^{a}\sigma \tilde{\nabla}_{b}\sigma \tilde{\nabla}^{b}\sigma + A_{35} \tilde{\nabla}^{a}\sigma \tilde{\nabla}_{b}\tilde{\nabla}_{a}P \tilde{\nabla}^{b}\sigma \nonumber \\ 
& + A_{36} \tilde{\nabla}^{a}P \tilde{\nabla}_{b}\tilde{\nabla}_{a}\sigma \tilde{\nabla}^{b}\sigma + A_{37} \tilde{\nabla}_{b}\tilde{\nabla}_{a}P \tilde{\nabla}^{b}\tilde{\nabla}^{a}P  + A_{38} \tilde{\nabla}_{b}\tilde{\nabla}_{a}\sigma \tilde{\nabla}^{b}\tilde{\nabla}^{a}\sigma\nonumber\\
\delta P^2_A=& A_{39} \tilde{R} \tilde{\nabla}_{a}\tilde{\nabla}^{a}\sigma + A_{40} \tilde{R} \tilde{\nabla}_{a}\sigma \tilde{\nabla}^{a}P + A_{41} \tilde{\nabla}_{a}\sigma \tilde{\nabla}^{a}\tilde{R} + A_{42} \tilde{R}^{ab} \tilde{\nabla}_{b}\tilde{\nabla}_{a}\sigma \nonumber \\ 
& + A_{43} \tilde{\nabla}_{a}\sigma \tilde{\nabla}^{a}P \tilde{\nabla}_{b}\tilde{\nabla}^{b}P + A_{44} \tilde{\nabla}_{a}\tilde{\nabla}^{a}P \tilde{\nabla}_{b}\tilde{\nabla}^{b}\sigma + A_{45} \tilde{\nabla}_{a}P \tilde{\nabla}^{a}P \tilde{\nabla}_{b}\tilde{\nabla}^{b}\sigma \nonumber \\ 
& + A_{46} \tilde{\nabla}_{a}\sigma \tilde{\nabla}^{a}\sigma \tilde{\nabla}_{b}\tilde{\nabla}^{b}\sigma + A_{47} \tilde{\nabla}^{a}\sigma \tilde{\nabla}_{b}\tilde{\nabla}^{b}\tilde{\nabla}_{a}P + A_{48} \tilde{\nabla}^{a}P \tilde{\nabla}_{b}\tilde{\nabla}^{b}\tilde{\nabla}_{a}\sigma \nonumber \\ 
& + A_{49} \tilde{\nabla}_{b}\tilde{\nabla}^{b}\tilde{\nabla}_{a}\tilde{\nabla}^{a}\sigma + A_{50} \tilde{\nabla}_{a}P \tilde{\nabla}^{a}P \tilde{\nabla}_{b}\sigma \tilde{\nabla}^{b}P + A_{51} \tilde{\nabla}^{a}P \tilde{\nabla}_{b}\tilde{\nabla}_{a}\sigma \tilde{\nabla}^{b}P \nonumber \\ 
& + A_{52} \tilde{R}_{ab} \tilde{\nabla}^{a}P \tilde{\nabla}^{b}\sigma + A_{53} \tilde{\nabla}_{a}\sigma \tilde{\nabla}^{a}P \tilde{\nabla}_{b}\sigma \tilde{\nabla}^{b}\sigma + A_{54} \tilde{\nabla}^{a}P \tilde{\nabla}_{b}\tilde{\nabla}_{a}P \tilde{\nabla}^{b}\sigma \nonumber \\ 
& + A_{55} \tilde{\nabla}^{a}\sigma \tilde{\nabla}_{b}\tilde{\nabla}_{a}\sigma \tilde{\nabla}^{b}\sigma + A_{56} \tilde{\nabla}_{b}\tilde{\nabla}_{a}\sigma \tilde{\nabla}^{b}\tilde{\nabla}^{a}P\nonumber\\
\delta g_{Aab}^2=&\tfrac{1}{2} A_{57}( \tilde{R} \tilde{\nabla}_{a}\sigma \tilde{\nabla}_{b}P + \tilde{R} \tilde{\nabla}_{a}P \tilde{\nabla}_{b}\sigma ) + \tfrac{1}{2} A_{58}( \tilde{\nabla}_{a}\sigma \tilde{\nabla}_{b}\tilde{R}+  \tilde{\nabla}_{a}\tilde{R} \tilde{\nabla}_{b}\sigma) \nonumber \\ 
& + A_{59} \tilde{R} \tilde{\nabla}_{b}\tilde{\nabla}_{a}\sigma + \tfrac{1}{2} A_{60} (\tilde{R}_{b}{}^{c} \tilde{\nabla}_{c}\tilde{\nabla}_{a}\sigma +  \tilde{R}_{a}{}^{c} \tilde{\nabla}_{c}\tilde{\nabla}_{b}\sigma )+ A_{61} \tilde{\nabla}_{b}\tilde{\nabla}_{a}\sigma \tilde{\nabla}_{c}\tilde{\nabla}^{c}P  \nonumber \\ 
&+\tfrac{1}{2} A_{62} (\tilde{\nabla}_{a}\sigma \tilde{\nabla}_{b}P \tilde{\nabla}_{c}\tilde{\nabla}^{c}P + \tilde{\nabla}_{a}P \tilde{\nabla}_{b}\sigma \tilde{\nabla}_{c}\tilde{\nabla}^{c}P ) + A_{63} \tilde{R}_{ab} \tilde{\nabla}_{c}\tilde{\nabla}^{c}\sigma  \nonumber \\ 
& + A_{64} \tilde{\nabla}_{a}P \tilde{\nabla}_{b}P \tilde{\nabla}_{c}\tilde{\nabla}^{c}\sigma + A_{65} \tilde{\nabla}_{a}\sigma \tilde{\nabla}_{b}\sigma \tilde{\nabla}_{c}\tilde{\nabla}^{c}\sigma + A_{66} \tilde{\nabla}_{b}\tilde{\nabla}_{a}P \tilde{\nabla}_{c}\tilde{\nabla}^{c}\sigma \nonumber \\ 
& + \tfrac{1}{2} A_{67}( \tilde{\nabla}_{b}\sigma \tilde{\nabla}_{c}\tilde{\nabla}^{c}\tilde{\nabla}_{a}P + \tilde{\nabla}_{a}\sigma \tilde{\nabla}_{c}\tilde{\nabla}^{c}\tilde{\nabla}_{b}P) + \tfrac{1}{2} A_{68} (\tilde{\nabla}_{b}P \tilde{\nabla}_{c}\tilde{\nabla}^{c}\tilde{\nabla}_{a}\sigma \nonumber \\ 
& + \tilde{\nabla}_{a}P \tilde{\nabla}_{c}\tilde{\nabla}^{c}\tilde{\nabla}_{b}\sigma) + A_{69} \tilde{\nabla}_{c}\tilde{\nabla}^{c}\tilde{\nabla}_{b}\tilde{\nabla}_{a}\sigma+ A_{70} \tilde{\nabla}_{b}\tilde{\nabla}_{a}\sigma \tilde{\nabla}_{c}P \tilde{\nabla}^{c}P + A_{71} \tilde{R}_{ab} \tilde{\nabla}_{c}\sigma \tilde{\nabla}^{c}P  \nonumber\\
& + A_{72} \tilde{\nabla}_{a}P \tilde{\nabla}_{b}P \tilde{\nabla}_{c}\sigma \tilde{\nabla}^{c}P + A_{73} \tilde{\nabla}_{a}\sigma \tilde{\nabla}_{b}\sigma \tilde{\nabla}_{c}\sigma \tilde{\nabla}^{c}P+ A_{74} \tilde{\nabla}_{b}\tilde{\nabla}_{a}P \tilde{\nabla}_{c}\sigma \tilde{\nabla}^{c}P  \nonumber \\ 
& + A_{75} \tilde{\nabla}_{c}\tilde{\nabla}_{b}\tilde{\nabla}_{a}\sigma \tilde{\nabla}^{c}P+ \tfrac{1}{2} A_{76}( \tilde{R}_{bc} \tilde{\nabla}_{a}\sigma \tilde{\nabla}^{c}P+  \tilde{R}_{ac} \tilde{\nabla}_{b}\sigma \tilde{\nabla}^{c}P )\nonumber \\ 
& + \tfrac{1}{2} A_{77} (\tilde{\nabla}_{a}\sigma \tilde{\nabla}_{b}P \tilde{\nabla}_{c}P \tilde{\nabla}^{c}P + \tilde{\nabla}_{a}P \tilde{\nabla}_{b}\sigma \tilde{\nabla}_{c}P \tilde{\nabla}^{c}P )\nonumber \\ 
& + \tfrac{1}{2} A_{78}( \tilde{\nabla}_{b}\sigma \tilde{\nabla}_{c}\tilde{\nabla}_{a}P \tilde{\nabla}^{c}P +  \tilde{\nabla}_{a}\sigma \tilde{\nabla}_{c}\tilde{\nabla}_{b}P \tilde{\nabla}^{c}P) \nonumber \\ 
& + \tfrac{1}{2} A_{79}( \tilde{\nabla}_{b}P \tilde{\nabla}_{c}\tilde{\nabla}_{a}\sigma \tilde{\nabla}^{c}P +  \tilde{\nabla}_{a}P \tilde{\nabla}_{c}\tilde{\nabla}_{b}\sigma \tilde{\nabla}^{c}P  ) \nonumber \\ 
&  + A_{80} \tilde{\nabla}_{c}\tilde{R}_{ab} \tilde{\nabla}^{c}\sigma+ A_{81} \tilde{\nabla}_{b}\tilde{\nabla}_{a}\sigma \tilde{\nabla}_{c}\sigma \tilde{\nabla}^{c}\sigma + A_{82} \tilde{\nabla}_{c}\tilde{\nabla}_{b}\tilde{\nabla}_{a}P \tilde{\nabla}^{c}\sigma \nonumber \\ 
&+ \tfrac{1}{2} A_{83} (\tilde{R}_{ac} \tilde{\nabla}_{b}P \tilde{\nabla}^{c}\sigma+ \tilde{R}_{bc} \tilde{\nabla}_{a}P \tilde{\nabla}^{c}\sigma)  + \tfrac{1}{2} A_{84}( \tilde{\nabla}_{b}\tilde{R}_{ac} \tilde{\nabla}^{c}\sigma  + \tilde{\nabla}_{a}\tilde{R}_{bc} \tilde{\nabla}^{c}\sigma ) \nonumber \\ 
&+ \tfrac{1}{2} A_{85} (\tilde{\nabla}_{a}\sigma \tilde{\nabla}_{b}P \tilde{\nabla}_{c}\sigma \tilde{\nabla}^{c}\sigma +  \tilde{\nabla}_{a}P \tilde{\nabla}_{b}\sigma \tilde{\nabla}_{c}\sigma \tilde{\nabla}^{c}\sigma )+ \tfrac{1}{2} A_{86}( \tilde{\nabla}_{b}P \tilde{\nabla}_{c}\tilde{\nabla}_{a}P \tilde{\nabla}^{c}\sigma \nonumber \\ 
& +  \tilde{\nabla}_{a}P \tilde{\nabla}_{c}\tilde{\nabla}_{b}P \tilde{\nabla}^{c}\sigma)+ \tfrac{1}{2} A_{87} (\tilde{\nabla}_{b}\sigma \tilde{\nabla}_{c}\tilde{\nabla}_{a}\sigma \tilde{\nabla}^{c}\sigma  +  \tilde{\nabla}_{a}\sigma \tilde{\nabla}_{c}\tilde{\nabla}_{b}\sigma \tilde{\nabla}^{c}\sigma) \nonumber \\ 
&  + \tfrac{1}{2} A_{88}( \tilde{\nabla}_{c}\tilde{\nabla}_{b}\sigma \tilde{\nabla}^{c}\tilde{\nabla}_{a}P +  \tilde{\nabla}_{c}\tilde{\nabla}_{a}\sigma \tilde{\nabla}^{c}\tilde{\nabla}_{b}P) + \tfrac{1}{2} A_{89} (\tilde{R}_{acbd} \tilde{\nabla}^{c}P \tilde{\nabla}^{d}\sigma+ \tilde{R}_{adbc} \tilde{\nabla}^{c}P \tilde{\nabla}^{d}\sigma )\nn\\
&+ A_{90} \tilde{R}_{acbd} \tilde{\nabla}^{d}\tilde{\nabla}^{c}\sigma+\tilde{g}_{ab}\Big( A_{91}  \tilde{R} \tilde{\nabla}_{c}\tilde{\nabla}^{c}\sigma + A_{92}  \tilde{R} \tilde{\nabla}_{c}\sigma \tilde{\nabla}^{c}P + A_{93}  \tilde{\nabla}_{c}\sigma \tilde{\nabla}^{c}\tilde{R}\nn\\
& + A_{94}  \tilde{R}^{cd} \tilde{\nabla}_{d}\tilde{\nabla}_{c}\sigma+ A_{95}  \tilde{\nabla}_{c}\sigma \tilde{\nabla}^{c}P \tilde{\nabla}_{d}\tilde{\nabla}^{d}P  + A_{96}  \tilde{\nabla}_{c}\tilde{\nabla}^{c}P \tilde{\nabla}_{d}\tilde{\nabla}^{d}\sigma  + A_{97}  \tilde{\nabla}_{c}P \tilde{\nabla}^{c}P \tilde{\nabla}_{d}\tilde{\nabla}^{d}\sigma\nn\\
&+ A_{98}  \tilde{\nabla}_{c}\sigma \tilde{\nabla}^{c}\sigma \tilde{\nabla}_{d}\tilde{\nabla}^{d}\sigma + A_{99}  \tilde{\nabla}^{c}\sigma \tilde{\nabla}_{d}\tilde{\nabla}^{d}\tilde{\nabla}_{c}P  + A_{100}  \tilde{\nabla}^{c}P \tilde{\nabla}_{d}\tilde{\nabla}^{d}\tilde{\nabla}_{c}\sigma + A_{101}  \tilde{\nabla}_{d}\tilde{\nabla}^{d}\tilde{\nabla}_{c}\tilde{\nabla}^{c}\sigma\nn\\
&+ A_{102}  \tilde{\nabla}_{c}P \tilde{\nabla}^{c}P \tilde{\nabla}_{d}\sigma \tilde{\nabla}^{d}P  + A_{103}  \tilde{\nabla}^{c}P \tilde{\nabla}_{d}\tilde{\nabla}_{c}\sigma \tilde{\nabla}^{d}P + A_{104}  \tilde{R}_{cd} \tilde{\nabla}^{c}P \tilde{\nabla}^{d}\sigma  \nn\\
& + A_{105}  \tilde{\nabla}_{c}\sigma \tilde{\nabla}^{c}P \tilde{\nabla}_{d}\sigma \tilde{\nabla}^{d}\sigma+ A_{106}  \tilde{\nabla}^{c}P \tilde{\nabla}_{d}\tilde{\nabla}_{c}P \tilde{\nabla}^{d}\sigma \nn\\
& + A_{107}  \tilde{\nabla}^{c}\sigma \tilde{\nabla}_{d}\tilde{\nabla}_{c}\sigma \tilde{\nabla}^{d}\sigma + A_{108}  \tilde{\nabla}_{d}\tilde{\nabla}_{c}\sigma \tilde{\nabla}^{d}\tilde{\nabla}^{c}P \Big)\labell{T2}
\end{align}
where $A_{11},\cdots A_{108}$ are some unknown coefficients. Some of the above terms  again correspond to the $d$-dimansional coordinate transformations. The presence of those terms does not affect our calculations because we are working in covariant approach, hence, we do not try to exclude them from the above list of corrections. 

The $ \mathbb{Z}_2 $-constraint for the higher order terms leaves the corrections in the class A to be  arbitrary and the corrections in class B to be written  in terms of corrections  in class A. In the next section we fix the arbitrary coefficients in the class A  by requiring the effective actions to be invariant under the above T-duality transformations.

 \section{  T-duality constraint on effective actions}

In the covariant approach for constructing the effective action from T-duality constraint, one   considers a D-dimensional covariant effective action $S_{\rm eff}$ which has the following $\alpha'$-expansion:
\beqa
S_{\rm eff}&=&\sum_{n=0}^{\infty}\alpha'^nS_n\labell{eff}
\eeqa
where $S_0$ is the    effective action at the leading order which   contains all covariant couplings at 2-derivative level with unknown coefficients, $S_1$ contains all covariant couplings at 4-derivative level with unknown coefficients, and so on. The invariance of the effective action \reef{eff} under the T-duality
means the following: One should reduce the theory on a circle with the killing direction $y$ to produce $d$-dimensional action $\bS_{\rm eff}$. Then one should transform this reduced action under the T-duality transformation \reef{T} to produce $\bS'_{\rm eff}$. The T-duality constraint is $\bS_{\rm eff}=\bS'_{\rm eff}$. In other words, 
\beqa
 \bS_{\rm eff}&\stackrel{T }{\longrightarrow}&\bS'_{\rm eff}=\bS_{\rm eff}\labell{Teff}
\eeqa
This constraint   is expected to be held at each order of $\alpha'$. That is , the action at the leading   order of  $\alpha'$ must be invariant under the Buscher rules, \ie 
\beqa
\bS_0&\stackrel{T^{(0)}}{\longrightarrow}&\bS'_0=\bS_0\,\labell{SS0}
\eeqa
At order $\alpha'$, the    invariance   means
\beqa
\bS_1& \xrightarrow{T^{(0)}}&\bS'_1=\bS_1-\delta \bS_1\,,\labell{deltaS}\\
\bS_0&\xrightarrow{T^{(0)}+\alpha'T^{(1)}} &\bS_0+\delta \bS_1+\cdots\nn
\eeqa
where $\delta\bS_1$ is at order $\alpha'$ and dots refer to the terms at higher orders of $\alpha'$ which are  produced by applying the T-duality transformation $T^{(1)}$ on   the reduced action $\bS_0$. The above relations mean that the effective action at order $\alpha'$ is not invariant under the Buscher rules, however, the non-invariance terms must be  reproduced by the transformation of the leading order action under the Buscher rules pluse their $\alpha'$-corrections.  

At order $(\alpha')^2$, the   invariance means  
\beqa
\bS_2& \xrightarrow{T^{(0)}} &\bS'_2=\bS_2- \delta \bS_2\,,\nonumber\\
\bS_0+\bS_1&\xrightarrow{T^{(0)}+\alpha'T^{(1)}+\alpha'^2T^{(2)}} &\bS_0+\bS_1+ \delta\bS_2 +\cdots\labell{deltaS1S2}
\eeqa
where $ \delta\bS_2$ is at order $\alpha'^2$. So if the effective action at order $\alpha'^2$ is not invariant under the Buscher rules, the non-invariance terms must be reproduced by the transformation of the lower order actions under the Buscher rules pulse $\alpha'$ and $\alpha'^2$ corrections.

 At order $(\alpha')^3$, the   invariance means  
\beqa
\bS_3& \xrightarrow{T^{(0)}} &\bS'_3=\bS_3-\delta \bS_3 \,,\nonumber\\
\bS_0+\bS_1+\bS_2&\xrightarrow{T^{(0)}+\alpha'T^{(1)}+\alpha'^2T^{(2)}+\alpha'^3T^{(3)}} &\bS_0+\bS_1+\bS_2 +\delta\bS_3  +\cdots\labell{s2s2}
\eeqa
where $\delta\bS_3 $ is at order $\alpha'^3$. Similarly for the   higher orders of $\alpha'$. Since  the constraint \reef{Teff} is on the effective actions, one is free to add   any total   derivative term to the $d$-dimensional Lagrangian in \reef{Teff}.

The corrections $\alpha' T^{(1)}+\alpha'^2T^{(2)}+\alpha'^3T^{(3)}+\cdots$ to the Buscher rule $T^{(0)}$, are in fact the higher derivative transformations to the $d$-dimensional fields, \ie
\beqa
\sigma&\rightarrow&-\sigma+\cdots\nn\\
P&\rightarrow &P+\cdots\nn\\
g_{ab}&\rightarrow &g_{ab}+\cdots\labell{df}
\eeqa
 As a result, the transformations in the second lines in \reef{deltaS}, \reef{deltaS1S2} and \reef{s2s2} are the transformations of $\bS_0$, $\bS_0+\bS_1$ and $\bS_0+\bS_1+\bS_2$, respectively, under the above field redefinitions. So, one may write  the T-duality constraint \reef{Teff} as
\beqa
 \bS_{\rm eff}&\stackrel{T^{(0)} }{\longrightarrow}&\bS'_{\rm eff}=\bS_{\rm eff}\,\,\, {\rm or }\,\,\, \delta\bS_{\rm eff}=0\labell{Teff0}
\eeqa
up to the $d$-dimensional field redefinitions \reef{df}.  

 The unknown coefficients in the reduced  action $\bS_{\rm eff}$ are inherited from the original action $S_{\rm eff}$, as a result, the constraint \reef{Teff} or \reef{Teff0} may fix the coefficients in  $S_{\rm eff} $. Let us check this for the trivial case at order $\alpha'^0$ in the next subsection.

\subsection{Effective action at order $\alpha'^0$}
We begin with the effective action at order $\alpha'^0$, which has 2-derivative couplings.  
The only 2-derivative couplings are:
\[
R\ ,\quad  \nabla_{\alpha}\Phi\nabla^{\alpha}\Phi \ , \quad\nabla_{\alpha}\nabla^{\alpha}\Phi
\]
However, using the fact that the effective action in the string frame has an overall factor of $e^{-2\Phi}\sqrt{-G}$, one realizes that the coupling $  \nabla_{\alpha}\nabla^{\alpha}\Phi  $ is related to $\nabla_{\alpha}\Phi\nabla^{\alpha}\Phi$ by using  integration by parts, so one can eliminate this coupling in the effective action. As a result, the most general covariant action at order $\alpha'^0$ is 
\beqa
S_0&=&-\frac{2}{\kappa^2}\int d^{d+1}x e^{-2\Phi}\sqrt{-G}\left(a_1R + a_2  \nabla_{\alpha}\Phi \nabla^{\alpha}\Phi\right)\labell{S0}
\eeqa
where $a_1,a_2$ are two unknown constants. 

The reduction of different terms in this action to $d$-dimensional spacetime is 
\begin{align}
e^{-2\Phi}\sqrt{-G}&= e^{-2P}\sqrt{-g} \nn\\
R&=\tilde{R} - 2 \tilde{\nabla}_{a}\tilde{\nabla}^{a}\sigma - 2 \tilde{\nabla}_{a}\sigma \tilde{\nabla}^{a}\sigma\nn\\
\nabla_{\alpha}\Phi \nabla^{\alpha}\Phi &=\tilde{\nabla}_{a}P \tilde{\nabla}^{a}P + \tilde{\nabla}_{a}\sigma \tilde{\nabla}^{a}P + \frac{1}{4} \tilde{\nabla}_{a}\sigma \tilde{\nabla}^{a}\sigma\labell{eq.1.70}
\end{align}
 where we have  assumed that the fields are independent of the killing coordinate $ y $. The reduction of the action \reef{S0} is then 
\begin{align}
 \bS_0&=-\frac{2}{\kappa^2}\int d^{d}x e^{-2P} \sqrt{-g}\Big(a_1\tilde{R}  + a_2 \tilde{\nabla}_{a}P \tilde{\nabla}^{a}P +(-4a_1+ a_2 )\tilde{\nabla}_{a}\sigma \tilde{\nabla}^{a}P\nn\\
&\qquad + \frac{1}{4} (-8a_1 + a_2) \tilde{\nabla}_{a}\sigma \tilde{\nabla}^{a}\sigma\Big)\labell{eq.1.80}
\end{align}
where we have also used the integration by parts in $d$-dimensional spacetime to write $\tilde{\nabla}_{a}\tilde{\nabla}^{a}\sigma=2\tilde{\nabla}_{a} P\tilde{\nabla}^{a}\sigma$. Now under the T-duality transformation  \reef{sigma}, it transforms to the following action:  
\begin{align}
\bS_0  \stackrel{T^{(0)}}{\longrightarrow}  \bS'_0&=-\frac{2}{\kappa^2}\int d^{d}x e^{-2P} \sqrt{-g}\Big(a_1\tilde{R}  + a_2 \tilde{\nabla}_{a}P \tilde{\nabla}^{a}P -(-4a_1+ a_2 )\tilde{\nabla}_{a}\sigma \tilde{\nabla}^{a}P\nn\\
&\qquad + \frac{1}{4} (-8a_1 + a_2) \tilde{\nabla}_{a}\sigma \tilde{\nabla}^{a}\sigma\Big)\labell{eq.1.100}
\end{align}
Requiring the two d-dimensionl actions to be identical, \ie \reef{SS0},  one finds  $ a_2=4a_1 $. This fixes the original D-dimensional action \reef{S0} up to an overall constant factor to be  
\begin{align}
S_0=-\frac{2}{\kappa^2}\int d^{d+1}x e^{-2\Phi}\sqrt{-G}\, a_1\left(R + 4\nabla_{\alpha}\Phi \nabla^{\alpha}\Phi\right)\,.\labell{eq.1.110}
\end{align}
This is the known effective action  at order $\alpha'$ in absence of $ B $-fields when the overall factor is $a_1=1$. In the next subsection, we continue the above calculations for the couplings at order $\alpha'$.

\subsection{Effective action at order $\alpha'$}

The covariant couplings at order $\alpha'$   have the following structures: 
\begin{align*}
&RR,\nabla\nabla R,\nabla\Phi\nabla R,\nabla\nabla\Phi R,\nabla\Phi\nabla\Phi R,\nabla\Phi\nabla\Phi\nabla\Phi\nabla\Phi,\nabla\nabla\Phi\nabla\nabla\Phi,\nabla\nabla\Phi\nabla\Phi\nabla\Phi,\\
&\nabla\Phi\nabla\nabla\nabla\Phi,\nabla\nabla\nabla\nabla\Phi
\end{align*}
where $R$ stands  for the Riemann curvature. One should consider all contractions in each structure. Using the Bianchi identities, one finds that there are only 16 independent couplings in the Lagrangian. 
However, in the action, one should consider all terms that are independent up to total derivative terms. In fact,   seven terms in the   Lagrangian are related to the other terms by some total derivative terms. They are  
\begin{align}
 \nabla_{\alpha}\nabla^{\alpha}R\ &\Rightarrow\   - 2R \nabla_{\alpha}\nabla^{\alpha}\Phi + 4R \nabla_{\alpha}\Phi \nabla^{\alpha}\Phi \nn\\
  \nabla_{\alpha}\Phi \nabla^{\alpha}R \ &\Rightarrow\  -R \nabla_{\alpha}\nabla^{\alpha}\Phi + 2R \nabla_{\alpha}\Phi \nabla^{\alpha}\Phi \nn\\
  R^{\alpha \beta} \nabla_{\beta}\nabla_{\alpha}\Phi\ &\Rightarrow\   \frac{1}{2}R \nabla_{\alpha}\nabla^{\alpha}\Phi - R \nabla_{\alpha}\Phi \nabla^{\alpha}\Phi  + 2R_{\alpha \beta} \nabla^{\alpha}\Phi \nabla^{\beta}\Phi \nn\\
 \nabla_{\beta}\nabla^{\beta}\nabla_{\alpha}\nabla^{\alpha}\Phi\ &\Rightarrow\   -2 \nabla_{\alpha}\nabla^{\alpha}\Phi \nabla_{\beta}\nabla^{\beta}\Phi + 4 \nabla_{\alpha}\Phi \nabla^{\alpha}\Phi \nabla_{\beta}\nabla^{\beta}\Phi  \nn\\
  \nabla^{\alpha}\Phi \nabla_{\beta}\nabla^{\beta}\nabla_{\alpha}\Phi \ &\Rightarrow\   - \nabla_{\alpha}\nabla^{\alpha}\Phi \nabla_{\beta}\nabla^{\beta}\Phi + 2 \nabla_{\alpha}\Phi \nabla^{\alpha}\Phi \nabla_{\beta}\nabla^{\beta}\Phi +R_{\alpha \beta} \nabla^{\alpha}\Phi \nabla^{\beta}\Phi \nn\\
 \nabla^{\alpha}\Phi \nabla_{\beta}\nabla_{\alpha}\Phi \nabla^{\beta}\Phi\ &\Rightarrow- \frac{1}{2} \nabla_{\alpha}\Phi \nabla^{\alpha}\Phi \nabla_{\beta}\nabla^{\beta}\Phi + \nabla_{\alpha}\Phi \nabla^{\alpha}\Phi \nabla_{\beta}\Phi \nabla^{\beta}\Phi \nn\\
  \nabla_{\beta}\nabla_{\alpha}\Phi \nabla^{\beta}\nabla^{\alpha}\Phi\ &\Rightarrow\   \nabla_{\alpha}\nabla^{\alpha}\Phi \nabla_{\beta}\nabla^{\beta}\Phi - 3 \nabla_{\alpha}\Phi \nabla^{\alpha}\Phi \nabla_{\beta}\nabla^{\beta}\Phi - R_{\alpha \beta} \nabla^{\alpha}\Phi \nabla^{\beta}\Phi \nn\\
 &\qquad\qquad\qquad+ 2 \nabla_{\alpha}\Phi \nabla^{\alpha}\Phi \nabla_{\beta}\Phi \nabla^{\beta}\Phi  \labell{eq.2.20}
\end{align}
where we have used the fact that the effective action   has the overall factor of $e^{-2\Phi}\sqrt{-G}$.  These seven identities, reduce the number of independent couplings  to  9 couplings, \ie  
\begin{align}
S_1=\frac{-2}{\kappa^2}\alpha'\int& d^{d+1}xe^{-2\Phi}\sqrt{-G}\Big(
b_1 R_{\alpha \beta \gamma \delta} R^{\alpha \beta \gamma \delta}+b_2 R_{\alpha \beta} R^{\alpha \beta} + b_3 R^2\nn\\ &+ b_4 R_{\alpha \beta} \nabla^{\alpha}\Phi \nabla^{\beta}\Phi+ b_5 R \nabla_{\alpha}\Phi \nabla^{\alpha}\Phi+b_6 R \nabla_{\alpha}\nabla^{\alpha}\Phi \nonumber \\ 
&  + b_7 \nabla_{\alpha}\nabla^{\alpha}\Phi \nabla_{\beta}\nabla^{\beta}\Phi + b_8 \nabla_{\alpha}\Phi \nabla^{\alpha}\Phi \nabla_{\beta}\nabla^{\beta}\Phi  + b_9 \nabla_{\alpha}\Phi \nabla^{\alpha}\Phi \nabla_{\beta}\Phi \nabla^{\beta}\Phi
\Big)\labell{eq.2.30}
\end{align}
where $b_1,\cdots, b_9$ are unknown coefficients. 

Apart from the coefficient of the Riemann squared term, all other 8 coefficients are a priori ambiguous because they are changed under field redefinition.
Consider transformation of effective action $ S_0 $ under the general field redefinitions $ G_{\mu\nu}\to G_{\mu\nu}+  \alpha'\delta G^{(1)}_{\mu\nu}+\cdots$ and $ \Phi\to \Phi+\alpha'\delta\Phi^{(1)}+\cdots $, \ie
\begin{align}
S_0\rightarrow S_0+
\alpha'\frac{\delta S_0}{\delta G_{\alpha\beta}} \delta G^{(1)}_{\alpha\beta}+
\alpha'\frac{\delta S_0}{\delta \Phi} \delta \Phi^{(1)}\labell{SS00}
\end{align}
The  variations at order $\alpha'$ are
\beqa
\delta G_{\mu\nu}^{(1)}&=&a_1R_{\mu\nu}+a_2\nabla_{\mu}\Phi\nabla_{\nu}\Phi+G_{\mu\nu}(a_3R+a_4\nabla_{\alpha}\Phi\nabla^{\alpha}\Phi+a_5\nabla_{\alpha}\nabla^{\alpha}\Phi)\nn\\
\delta \Phi^{(1)}&=&c_1R+c_2\nabla_{\alpha}\Phi\nabla^{\alpha}\Phi+c_3\nabla_{\alpha}\nabla^{\alpha}\Phi\labell{dd}
\eeqa
None of them correspond to the coordinate transformations.  Under the above transformation, all the $ b_i $-coefficients (except $ b_1 $) in \reef{eq.2.30} are changed to $ b'_i $ such that one particular combination of these terms is invariant \cite{Metsaev:1987zx}, \ie
 \beqa
\delta b_9+2\delta b_8+4 \delta b_7-8 \delta b_6-4 \delta b_5+16 \delta b_3=0\labell{fcon}
\eeqa
where $\delta b_i=b'_i-b_i$. In other words, if one defines 
 \begin{align}
 \xi\equiv b_9+2 b_8+4  b_7-8 b_6-4  b_5+16  b_3 \labell{eq.xi}
\end{align} 
Then $\xi$ is  invariant under field redefinitions.  

From the transformation of $S_0$ under the field redefinitions, \ie \reef{SS00}, one realizes that the effect of field redefinitions in the action at order $\alpha'$  is the same as   all possible  substitutions of the equations of motion at order $\alpha'^0$, \ie $\frac{\delta S_0}{\delta G_{\alpha\beta}} =\frac{\delta S_0}{\delta \Phi}=0$, in the effective action at order $\alpha'$. Hence, if one is not interested in the explicit form of the field redefinitions,  one may use equations of motion
to  remove the ambiguous coefficients in the effective action \reef{eq.2.30}. In fact using 
\beqa
R_{\alpha\beta}+2\nabla_{\alpha}\nabla_{\beta}\Phi&=&0\nn\\
R+4\nabla_{\alpha}\Phi\nabla^{\alpha}\Phi&=&0\labell{eom}
\eeqa
one can write the action \reef{eq.2.30} as 
\beqa
S_1=\frac{-2}{\kappa^2}\alpha'\int& d^{d+1}xe^{-2\Phi}\sqrt{-G}\Big(
b_1 R_{\alpha \beta \gamma \delta} R^{\alpha \beta \gamma \delta}+   \xi\nabla_{\alpha}\Phi \nabla^{\alpha}\Phi \nabla_{\beta}\Phi \nabla^{\beta}\Phi
\Big)\labell{S111}
\eeqa
The last term may be written in other forms using the equations of motion \reef{eom}. One can use the field redefinitions \reef{dd} to rewrite \reef{eq.2.30} as \reef{S111}. In performing this calculation, one finds one of the coefficients $a_3,a_4,a_5,c_1,c_2,c_3$ to be arbitrary. We will find that similar calculations in the $d$-dimensional theory leave two coefficients in the $d$-dimensional field redefinitions \reef{eq.2.2.20} to be arbitrary.

Since $\xi$ is invariant under the field redefinition, one can find this function   from the S-matrix calculation. The field redefinition allows us to choose seven coefficients among the coefficients $b_2,\cdots b_9$ to be arbitrary, so one may choose   $b_2=b_3=b_4=b_5=b_6=b_7=b_8=0$. Then $\xi=b_9$. For this choice for the coefficients, on the other hand,  the S-matrix calculation fixes $b_9=0$ \cite{Metsaev:1986yb}, so $\xi=0$. Therefore, the effective action \reef{finalS10} is consistent with S-matrix for any value for coefficients $b_2,\cdots, b_8$. 

We are going, however,  to find this function  from the T-duality constraint \reef{deltaS}.   To this end, one first needs to reduce the effective    action \reef{eq.2.30} to the $d$-dimensional spacetime, \ie
\begin{align}
\bS_1=&-\frac{2}{\kappa^2}\alpha'\int  d^dx e^{-2P}\sqrt{-g}\Big(
 b_1 \tilde{R}_{abcd} \tilde{R}^{abcd}+b_2 \tilde{R}_{ab} \tilde{R}^{ab} + b_3 \tilde{R}^2  + b_6 \tilde{R} \tilde{\nabla}_{a}\tilde{\nabla}^{a}P  \nonumber \\ 
& + b_5 \tilde{R} \tilde{\nabla}_{a}P \tilde{\nabla}^{a}P + (b_5 + b_6) \tilde{R} \tilde{\nabla}_{a}\sigma \tilde{\nabla}^{a}P + \tfrac{1}{4} (-16 b_3 + b_5 + 2 b_6) \tilde{R} \tilde{\nabla}_{a}\sigma \tilde{\nabla}^{a}\sigma \nonumber \\ 
& - 2 b_2 \tilde{R}^{ab} \tilde{\nabla}_{b}\tilde{\nabla}_{a}\sigma + b_7 \tilde{\nabla}_{a}\tilde{\nabla}^{a}P \tilde{\nabla}_{b}\tilde{\nabla}^{b}P + b_8 \tilde{\nabla}_{a}P \tilde{\nabla}^{a}P \tilde{\nabla}_{b}\tilde{\nabla}^{b}P \nonumber \\ 
&+ (2 b_7  + b_8) \tilde{\nabla}_{a}\sigma \tilde{\nabla}^{a}P \tilde{\nabla}_{b}\tilde{\nabla}^{b}P + (-2 b_6 + b_7 + \tfrac{1}{4} b_8) \tilde{\nabla}_{a}\sigma \tilde{\nabla}^{a}\sigma \tilde{\nabla}_{b}\tilde{\nabla}^{b}P \nonumber \\ 
&  + (-2 b_6+ b_7) \tilde{\nabla}_{a}\tilde{\nabla}^{a}P \tilde{\nabla}_{b}\tilde{\nabla}^{b}\sigma + (b_2 + 4 b_3 -  b_6 + \tfrac{1}{4} b_7) \tilde{\nabla}_{a}\tilde{\nabla}^{a}\sigma \tilde{\nabla}_{b}\tilde{\nabla}^{b}\sigma \nonumber \\ 
& + \tfrac{1}{2} (-4 b_5 + b_8) \tilde{\nabla}_{a}P \tilde{\nabla}^{a}P \tilde{\nabla}_{b}\tilde{\nabla}^{b}\sigma + (-2 b_5 - 2 b_6 + b_7 + \tfrac{1}{2} b_8) \tilde{\nabla}_{a}\sigma \tilde{\nabla}^{a}P \tilde{\nabla}_{b}\tilde{\nabla}^{b}\sigma \nonumber \\ 
& + \tfrac{1}{8} (16 b_2 + 64 b_3 - 4 b_5 - 16 b_6 + 4 b_7 + b_8) \tilde{\nabla}_{a}\sigma \tilde{\nabla}^{a}\sigma \tilde{\nabla}_{b}\tilde{\nabla}^{b}\sigma + b_4 \tilde{R}_{ab} \tilde{\nabla}^{a}P \tilde{\nabla}^{b}P \nonumber \\ 
& + b_9 \tilde{\nabla}_{a}P \tilde{\nabla}^{a}P \tilde{\nabla}_{b}P \tilde{\nabla}^{b}P + (b_8 + 2 b_9) \tilde{\nabla}_{a}P \tilde{\nabla}^{a}P \tilde{\nabla}_{b}\sigma \tilde{\nabla}^{b}P  \nonumber \\ 
&+ (- b_4 + b_7 + b_8 + b_9) \tilde{\nabla}_{a}\sigma \tilde{\nabla}^{a}P \tilde{\nabla}_{b}\sigma \tilde{\nabla}^{b}P -  b_4 \tilde{\nabla}^{a}P \tilde{\nabla}_{b}\tilde{\nabla}_{a}\sigma \tilde{\nabla}^{b}P + b_4 \tilde{R}_{ab} \tilde{\nabla}^{a}P \tilde{\nabla}^{b}\sigma \nonumber \\ 
& + \tfrac{1}{4} (-8 b_2 + b_4) \tilde{R}_{ab} \tilde{\nabla}^{a}\sigma \tilde{\nabla}^{b}\sigma + \tfrac{1}{2} (-4 b_5 + b_8 + b_9) \tilde{\nabla}_{a}P \tilde{\nabla}^{a}P \tilde{\nabla}_{b}\sigma \tilde{\nabla}^{b}\sigma \nonumber \\ 
& + (- b_4 - 2 b_5 - 2 b_6 + b_7 + \tfrac{3}{4} b_8 + \tfrac{1}{2} b_9) \tilde{\nabla}_{a}\sigma \tilde{\nabla}^{a}P \tilde{\nabla}_{b}\sigma \tilde{\nabla}^{b}\sigma  \nonumber \\ 
&+ \tfrac{1}{16} (64 b_1 + 32 b_2 + 64 b_3 - 4 b_4 - 8 b_5 - 16 b_6 + 4 b_7 + 2 b_8 + b_9) \tilde{\nabla}_{a}\sigma \tilde{\nabla}^{a}\sigma \tilde{\nabla}_{b}\sigma \tilde{\nabla}^{b}\sigma  \nonumber \\ 
&-  b_4 \tilde{\nabla}^{a}P \tilde{\nabla}_{b}\tilde{\nabla}_{a}\sigma \tilde{\nabla}^{b}\sigma + (8 b_1 + 2 b_2 -  \tfrac{1}{4} b_4) \tilde{\nabla}^{a}\sigma \tilde{\nabla}_{b}\tilde{\nabla}_{a}\sigma \tilde{\nabla}^{b}\sigma  \nonumber \\ 
& + \tfrac{1}{2} (-8 b_3 + b_6) \tilde{R} \tilde{\nabla}_{a}\tilde{\nabla}^{a}\sigma+ (4 b_1+ b_2) \tilde{\nabla}_{b}\tilde{\nabla}_{a}\sigma \tilde{\nabla}^{b}\tilde{\nabla}^{a}\sigma
 \Big)\labell{eq.2.35}
\end{align}
Using  the Buscher rule \reef{sigma}, one can easily find the $d$-dimensional dual action. It can be written as  $ \bS_1'=\bS_1-\delta\bS_1$ where $\delta\bS_1$ is   twice the terms in \reef{eq.2.35} that have odd number of $\sigma$, \ie
\begin{align}
\delta\bS_1=& \int  d^dx e^{-2P}\sqrt{-g}\Big(
   (b_5 + b_6) \tilde{R} \tilde{\nabla}_{a}\sigma \tilde{\nabla}^{a}P  - 2 b_2 \tilde{R}^{ab} \tilde{\nabla}_{b}\tilde{\nabla}_{a}\sigma  + (2 b_7  + b_8) \tilde{\nabla}_{a}\sigma \tilde{\nabla}^{a}P \tilde{\nabla}_{b}\tilde{\nabla}^{b}P    \nonumber \\ 
&  + (-2 b_6+ b_7) \tilde{\nabla}_{a}\tilde{\nabla}^{a}P \tilde{\nabla}_{b}\tilde{\nabla}^{b}\sigma+ \tfrac{1}{2} (-4 b_5 + b_8) \tilde{\nabla}_{a}P \tilde{\nabla}^{a}P \tilde{\nabla}_{b}\tilde{\nabla}^{b}\sigma + \tfrac{1}{2} (-8 b_3 + b_6) \tilde{R} \tilde{\nabla}_{a}\tilde{\nabla}^{a}\sigma  \nonumber \\ 
& + \tfrac{1}{8} (16 b_2 + 64 b_3 - 4 b_5 - 16 b_6 + 4 b_7 + b_8) \tilde{\nabla}_{a}\sigma \tilde{\nabla}^{a}\sigma \tilde{\nabla}_{b}\tilde{\nabla}^{b}\sigma  + (b_8 + 2 b_9) \tilde{\nabla}_{a}P \tilde{\nabla}^{a}P \tilde{\nabla}_{b}\sigma \tilde{\nabla}^{b}P  \nn\\
& + (8 b_1 + 2 b_2 -  \tfrac{1}{4} b_4) \tilde{\nabla}^{a}\sigma \tilde{\nabla}_{b}\tilde{\nabla}_{a}\sigma \tilde{\nabla}^{b}\sigma    -  b_4 \tilde{\nabla}^{a}P \tilde{\nabla}_{b}\tilde{\nabla}_{a}\sigma \tilde{\nabla}^{b}P + b_4 \tilde{R}_{ab} \tilde{\nabla}^{a}P \tilde{\nabla}^{b}\sigma  \nonumber \\ 
&  + (- b_4 - 2 b_5 - 2 b_6 + b_7 + \tfrac{3}{4} b_8 + \tfrac{1}{2} b_9) \tilde{\nabla}_{a}\sigma \tilde{\nabla}^{a}P \tilde{\nabla}_{b}\sigma \tilde{\nabla}^{b}\sigma  
 \Big)\left(-\frac{4}{\kappa^2}\alpha'\right) \labell{delS}
\end{align} 
This should be zero up to field redefinitions.

On the other hand, the $d$-dimensional effective action at the leading order of $\alpha'$ that we have found in the previous subsection  is 
\beqa
\bS_0&=&-\frac{2}{\kappa^2}\int d^{d}x e^{-2P} \sqrt{-g}\, \Big(\tilde{R}  + 4 \tilde{\nabla}_{a}P \tilde{\nabla}^{a}P  - \tilde{\nabla}_{a}\sigma \tilde{\nabla}^{a}\sigma\Big)\labell{LS}
\eeqa
Variation of this action under the T-duality transformation \reef{eq.2.2.20} at order $\alpha'$ is
\begin{align}
{  \delta \bS_1}=&-\alpha'\frac{2}{\kappa^2}\int d^{d}x e^{-2P} \sqrt{-g}\Big( -(\tilde{R}^{ab}  +2 \tilde{\nabla}^{a} \tilde{\nabla}^{b}P -\tilde{\nabla}^{a}\sigma \tilde{\nabla}^{b}\sigma\nonumber\\
&-\tfrac{1}{2}\tilde{g}^{ab}  (\tilde{R}+  4\tilde{\nabla}_{c} \tilde{\nabla}^{c}P   - 4    \tilde{\nabla}_{c}P \tilde{\nabla}^{c}P-  
 \tilde{\nabla}_{c}\sigma \tilde{\nabla}^{c}\sigma)\Big)\delta g^{1}{}_{Aab} \nonumber \\ 
& - 2 (\tilde{R}+4 \tilde{\nabla}_{a}\tilde{\nabla}^{a}P -4 \tilde{\nabla}_{a}P \tilde{\nabla}^{a}P - \tilde{\nabla}_{a}\sigma \tilde{\nabla}^{a}\sigma)\delta P^1_A-2( \tilde{\nabla}_{a}\tilde{\nabla}^{a}\sigma - 2 \tilde{\nabla}_{a}\sigma \tilde{\nabla}^{a}P)\delta\sigma^1_A  \Big)\labell{eq.2.2.10}
\end{align}
where $\delta \sigma_A^{1},\ \delta P_A^{1},  \delta g^{1}{}_{Aab}$ are given in \reef{eq.2.2.20}.

According to \eqref{deltaS}, by equating the right hand sides of the equations \reef{delS} and \reef{eq.2.2.10}, one finds the following constraint:
\beqa
&&-\alpha'\frac{2}{\kappa^2}\int d^{d}x e^{-2P} \sqrt{-g}\Big[
\tfrac{1}{2} (4 A_5 -  A_8 + 3 A_9 - 16 b_3 + 2 b_6 -  A_9 D) \tilde{R} \tilde{\nabla}_{a}\tilde{\nabla}^{a}\sigma\nonumber\\
&& + \tfrac{1}{2} \bigl(4 A_6 -  A_7 + 4 b_5 + 4 b_6 -  A_{10} (-3  + D)\bigr) \tilde{R} \tilde{\nabla}_{a}\sigma \tilde{\nabla}^{a}P -  \tfrac{1}{2} (4 A_1 + A_8) \tilde{\nabla}_{a}\sigma \tilde{\nabla}^{a}\tilde{R}\nonumber\\
&& - 4 b_2 \tilde{R}^{ab} \tilde{\nabla}_{b}\tilde{\nabla}_{a}\sigma + 2 (2 b_7 + b_8) \tilde{\nabla}_{a}\sigma \tilde{\nabla}^{a}P \tilde{\nabla}_{b}\tilde{\nabla}^{b}P -  (A_7 + 4 b_6 - 2 b_7) \tilde{\nabla}_{a}\tilde{\nabla}^{a}P \tilde{\nabla}_{b}\tilde{\nabla}^{b}\sigma\nonumber\\
&& + (8 A_5 - 2 A_8 + 6 A_9  - 4 b_5 + b_8 - 2 A_9 D) \tilde{\nabla}_{a}P \tilde{\nabla}^{a}P \tilde{\nabla}_{b}\tilde{\nabla}^{b}\sigma\nonumber\\
&& + \tfrac{1}{4} (-8 A_5 + 2 A_8 - 6 A_9 + 16 b_2 + 64 b_3 - 4 b_5 - 16 b_6 + 4 b_7  + b_8 + 2 A_9 D) \tilde{\nabla}_{a}\sigma \tilde{\nabla}^{a}\sigma \tilde{\nabla}_{b}\tilde{\nabla}^{b}\sigma \nonumber\\
&&+ \bigl(-2 A_2 + A_{10} (-2 + D)\bigr) \tilde{\nabla}^{a}\sigma \tilde{\nabla}_{b}\tilde{\nabla}^{b}\tilde{\nabla}_{a}P + \bigl(-8 A_5  + A_{10} (-2 + D)\bigr) \tilde{\nabla}^{a}P \tilde{\nabla}_{b}\tilde{\nabla}^{b}\tilde{\nabla}_{a}\sigma\nonumber\\
&& + A_9 (-2 + D) \tilde{\nabla}_{b}\tilde{\nabla}^{b}\tilde{\nabla}_{a}\tilde{\nabla}^{a}\sigma + 2 \bigl(4 A_6 + A_7 + b_8 + 2 b_9  -  A_{10} (-3 + D)\bigr) \tilde{\nabla}_{a}P \tilde{\nabla}^{a}P \tilde{\nabla}_{b}\sigma \tilde{\nabla}^{b}P\nonumber\\
&& - 2 (4 A_6 - 2 A_8 + b_4) \tilde{\nabla}^{a}P \tilde{\nabla}_{b}\tilde{\nabla}_{a}\sigma \tilde{\nabla}^{b}P + 2 (A_2 + 4 A_5  + A_7 + b_4) \tilde{R}_{ab} \tilde{\nabla}^{a}P \tilde{\nabla}^{b}\sigma\nonumber\\
&& -  \tfrac{1}{2} \bigl(4 A_6 + A_7 + 4 b_4 + 8 b_5 + 8 b_6 - 4 b_7 - 3 b_8 - 2 b_9 -  A_{10} (-3 + D)\bigr) \tilde{\nabla}_{a}\sigma \tilde{\nabla}^{a}P \tilde{\nabla}_{b}\sigma \tilde{\nabla}^{b}\sigma\nonumber\\
&& - 4 (A_3 + 2 A_6) \tilde{\nabla}^{a}P \tilde{\nabla}_{b}\tilde{\nabla}_{a}P \tilde{\nabla}^{b}\sigma -  \tfrac{1}{2} (8 A_4 + 2 A_8 - 32 b_1  - 8 b_2 + b_4) \tilde{\nabla}^{a}\sigma \tilde{\nabla}_{b}\tilde{\nabla}_{a}\sigma \tilde{\nabla}^{b}\sigma\nonumber\\
&& + \bigl(A_7 + 2 A_{10} (-2 + D)\bigr) \tilde{\nabla}_{b}\tilde{\nabla}_{a}\sigma \tilde{\nabla}^{b}\tilde{\nabla}^{a}P
 \Big] =0\labell{con}
\eeqa
Not all the above 18 terms are independent. One should subtract total derivative terms to find independent constraints. There are nine $d$-dimensional total derivative terms, \ie 
\begin{align}
\tilde{R}^{ab} \tilde{\nabla}_{b}\tilde{\nabla}_{a}\sigma &=\tfrac{1}{2} \tilde{R} \tilde{\nabla}_{a}\tilde{\nabla}^{a}\sigma -  \tilde{R} \tilde{\nabla}_{a}\sigma \tilde{\nabla}^{a}P + 2 \tilde{R}_{ab} \tilde{\nabla}^{a}P \tilde{\nabla}^{b}\sigma\nn\\
\tilde{\nabla}_{a}\sigma \tilde{\nabla}^{a}\tilde{R}&=- \tilde{R} \tilde{\nabla}_{a}\tilde{\nabla}^{a}\sigma + 2 \tilde{R} \tilde{\nabla}_{a}\sigma \tilde{\nabla}^{a}P\nn\\
\tilde{\nabla}^{a}\sigma \tilde{\nabla}_{b}\tilde{\nabla}_{a}\sigma \tilde{\nabla}^{b}\sigma&=- \tfrac{1}{2} \tilde{\nabla}_{a}\sigma \tilde{\nabla}^{a}\sigma \tilde{\nabla}_{b}\tilde{\nabla}^{b}\sigma + \tilde{\nabla}_{a}\sigma \tilde{\nabla}^{a}P \tilde{\nabla}_{b}\sigma \tilde{\nabla}^{b}\sigma\nn\\
\tilde{\nabla}_{b}\tilde{\nabla}^{b}\tilde{\nabla}_{a}\tilde{\nabla}^{a}\sigma&=-2 \tilde{\nabla}_{a}\tilde{\nabla}^{a}P \tilde{\nabla}_{b}\tilde{\nabla}^{b}\sigma + 4 \tilde{\nabla}_{a}P \tilde{\nabla}^{a}P \tilde{\nabla}_{b}\tilde{\nabla}^{b}\sigma\nn\\
\tilde{\nabla}^{a}P \tilde{\nabla}_{b}\tilde{\nabla}^{b}\tilde{\nabla}_{a}\sigma&=- \tilde{\nabla}_{a}\tilde{\nabla}^{a}P \tilde{\nabla}_{b}\tilde{\nabla}^{b}\sigma + 2 \tilde{\nabla}_{a}P \tilde{\nabla}^{a}P \tilde{\nabla}_{b}\tilde{\nabla}^{b}\sigma + \tilde{R}_{ab} \tilde{\nabla}^{a}P \tilde{\nabla}^{b}\sigma\nn\\
\tilde{\nabla}^{a}P \tilde{\nabla}_{b}\tilde{\nabla}_{a}\sigma \tilde{\nabla}^{b}P&=- \tilde{\nabla}_{a}\sigma \tilde{\nabla}^{a}P \tilde{\nabla}_{b}\tilde{\nabla}^{b}P + \tfrac{1}{2} \tilde{\nabla}_{a}P \tilde{\nabla}^{a}P \tilde{\nabla}_{b}\tilde{\nabla}^{b}\sigma + \tilde{\nabla}_{a}P \tilde{\nabla}^{a}P \tilde{\nabla}_{b}\sigma \tilde{\nabla}^{b}P\nn\\
\tilde{\nabla}^{a}P \tilde{\nabla}_{b}\tilde{\nabla}_{a}P \tilde{\nabla}^{b}\sigma&=- \tfrac{1}{2} \tilde{\nabla}_{a}P \tilde{\nabla}^{a}P \tilde{\nabla}_{b}\tilde{\nabla}^{b}\sigma + \tilde{\nabla}_{a}P \tilde{\nabla}^{a}P \tilde{\nabla}_{b}\sigma \tilde{\nabla}^{b}P\nn\\
\tilde{\nabla}^{a}\sigma \tilde{\nabla}_{b}\tilde{\nabla}^{b}\tilde{\nabla}_{a}P&=2 \tilde{\nabla}_{a}\sigma \tilde{\nabla}^{a}P \tilde{\nabla}_{b}\tilde{\nabla}^{b}P -  \tilde{\nabla}_{a}\tilde{\nabla}^{a}P \tilde{\nabla}_{b}\tilde{\nabla}^{b}\sigma + \tilde{R}_{ab} \tilde{\nabla}^{a}P \tilde{\nabla}^{b}\sigma\nn\\
\tilde{\nabla}_{b}\tilde{\nabla}_{a}\sigma \tilde{\nabla}^{b}\tilde{\nabla}^{a}P&=-2 \tilde{\nabla}_{a}\sigma \tilde{\nabla}^{a}P \tilde{\nabla}_{b}\tilde{\nabla}^{b}P + \tilde{\nabla}_{a}\tilde{\nabla}^{a}P \tilde{\nabla}_{b}\tilde{\nabla}^{b}\sigma -  \tilde{\nabla}_{a}P \tilde{\nabla}^{a}P \tilde{\nabla}_{b}\tilde{\nabla}^{b}\sigma \nn\\
&\qquad+ 2 \tilde{\nabla}_{a}P \tilde{\nabla}^{a}P \tilde{\nabla}_{b}\sigma \tilde{\nabla}^{b}P -  \tilde{R}_{ab} \tilde{\nabla}^{a}P \tilde{\nabla}^{b}\sigma\labell{tot}
\end{align}
Using the above total derivative terms, one finds the following constraint in which all terms are independent:
\begin{align}
&\int d^{d}x e^{-2P} \sqrt{-g}\Big[
\tfrac{1}{2} (4 A_1^{\text{}} + 4 A_5^{\text{}} + 3 A_9^{\text{}} - 4 b_2^{\text{}} - 16 b_3^{\text{}} + 2 b_6^{\text{}} -  A_9^{\text{}} D) \tilde{R} \tilde{\nabla}_{a}\tilde{\nabla}^{a}\sigma\nn\\
& -  \tfrac{1}{2} \bigl(8 A_1^{\text{}} - 4 A_6^{\text{}} + A_7^{\text{}} + 2 A_8^{\text{}} - 8 b_2^{\text{}} - 4 b_5^{\text{}} - 4 b_6^{\text{}} + A_{10}^{\text{}} (-3 + D)\bigr) \tilde{R} \tilde{\nabla}_{a}\sigma \tilde{\nabla}^{a}P \nn\\
&+ 2 \bigl(-2 A_2^{\text{}} + 4 A_6^{\text{}} -  A_7^{\text{}} - 2 A_8^{\text{}} + b_4^{\text{}} + 2 b_7^{\text{}} + b_8^{\text{}} -  A_{10}^{\text{}} (-2 + D)\bigr) \tilde{\nabla}_{a}\sigma \tilde{\nabla}^{a}P \tilde{\nabla}_{b}\tilde{\nabla}^{b}P\nn\\
& + 2 (A_2^{\text{}} + 4 A_5^{\text{}} + 2 A_9^{\text{}} - 2 b_6^{\text{}} + b_7^{\text{}} -  A_9^{\text{}} D) \tilde{\nabla}_{a}\tilde{\nabla}^{a}P \tilde{\nabla}_{b}\tilde{\nabla}^{b}\sigma \nn\\
&+ (2 A_3^{\text{}} - 8 A_5^{\text{}} -  A_7^{\text{}} - 2 A_9^{\text{}} -  b_4^{\text{}} - 4 b_5^{\text{}} + b_8^{\text{}} + 2 A_9^{\text{}} D) \tilde{\nabla}_{a}P \tilde{\nabla}^{a}P \tilde{\nabla}_{b}\tilde{\nabla}^{b}\sigma \nn\\
&+ \tfrac{1}{4} (8 A_4^{\text{}} - 8 A_5^{\text{}} + 4 A_8^{\text{}} - 6 A_9^{\text{}} - 32 b_1^{\text{}} + 8 b_2^{\text{}} + 64 b_3^{\text{}} + b_4^{\text{}} - 4 b_5^{\text{}} - 16 b_6^{\text{}}\nn\\
& \qquad + 4 b_7^{\text{}} + b_8^{\text{}}+ 2 A_9^{\text{}} D) \tilde{\nabla}_{a}\sigma \tilde{\nabla}^{a}\sigma \tilde{\nabla}_{b}\tilde{\nabla}^{b}\sigma\nn\\
& + 2 \bigl(-2 A_3^{\text{}} - 4 A_6^{\text{}} + 2 A_7^{\text{}} + 2 A_8^{\text{}} -  b_4^{\text{}} + b_8^{\text{}} + 2 b_9^{\text{}} + A_{10}^{\text{}} (-1 + D)\bigr) \tilde{\nabla}_{a}P \tilde{\nabla}^{a}P \tilde{\nabla}_{b}\sigma \tilde{\nabla}^{b}P\nn\\
& + (A_7^{\text{}} - 8 b_2^{\text{}} + 2 b_4^{\text{}}) \tilde{R}_{ab} \tilde{\nabla}^{a}P \tilde{\nabla}^{b}\sigma \nn\\
&-  \tfrac{1}{2} \bigl(8 A_4^{\text{}} + 4 A_6^{\text{}} + A_7^{\text{}} + 2 A_8^{\text{}} - 32 b_1^{\text{}} - 8 b_2^{\text{}} + 5 b_4^{\text{}} + 8 b_5^{\text{}} + 8 b_6^{\text{}} - 4 b_7^{\text{}} - 3 b_8^{\text{}} - 2 b_9^{\text{}} \nn\\
&\qquad-  A_{10}^{\text{}} (-3 + D)\bigr) \tilde{\nabla}_{a}\sigma \tilde{\nabla}^{a}P \tilde{\nabla}_{b}\sigma \tilde{\nabla}^{b}\sigma
 \Big] =0
 \labell{con1}
\end{align}
Solving the above 9 independent constraints, one finds effective action at order $\alpha'$ and its corresponding T-duality transformation. 

One finds the following solution for the constraints \reef{con1}:
\begin{align}
& A_{1} =  \tfrac{1}{8} ( 4 A_6 - 2 A_8- A_{10}(D-3) + 2 b_4 + 4 b_5 + 4 b_6 ), \nonumber\\
&   A_{2} = \tfrac{1}{2} \bigl(4 A_6 - 2 A_8  -  A_{10} ( D-2)- 8 b_2 + 3 b_4 + 2 b_7 + b_8\bigr), \nonumber\\
& A_3=\tfrac{1}{2} \bigl(-4 A_6 + 2 A_8+ A_{10} ( D-1) + 16 b_2 - 32 b_3 - 5 b_4 + 8 b_5 + 16 b_6 - 8 b_7 - 3 b_8 \bigr),\nonumber\\
& A_{4}= \tfrac{1}{8} \bigl(-4 A_6 - 2 A_8+ A_{10} ( D-3) + 32 b_1 - 32 b_3 - 3 b_4 + 8 b_6 - 4 b_7 -  b_8 \bigr), \nonumber\\
& A_{5} =  \tfrac{1}{8} (-4 A_6 + 2 A_8 + 8 b_2 +32(D-2) b_3 +(D- 5) b_4 -4(D-2) b_5 - 4(3D-7) b_6 \nn\\
&\qquad+4(D- 3) b_7 +(D- 3) b_8 ), \nonumber\\
&  A_{7} =  8 b_2 - 2 b_4, \nonumber\\
& A_{9}=\tfrac{1}{2} (- A_{10} + 32 b_3 + b_4 - 4 b_5 - 12 b_6 + 4 b_7 + b_8),\nonumber\\
& { \xi}= 0\labell{c1}
\end{align}
where $\xi$ is the combination of the coefficients in \reef{eq.2.30} that appears in \reef{eq.xi}. The last equation in \reef{c1} gives exactly the value for $\xi$ which is fixed by the S-matrix calculation. Using this value for $\xi$, one finds the T-duality invariant effective action at order $\alpha'$ to be \reef{finalS10} for the arbitrary coefficients $b_1,b_2,\cdots, b_9$.  
The corresponding T-duality transformations  are
\begin{align}
\delta\sigma^{1}_A=&\tfrac{1}{8} ( 4 A_6 - 2 A_8- A_{10}(D-3) + 2 b_4 + 4 b_5 + 4 b_6 ) \tilde{R} \nn\\
&+ \tfrac{1}{2} \bigl(4 A_6 - 2 A_8  -  A_{10} ( D-2)- 8 b_2 + 3 b_4 + 2 b_7 + b_8\bigr) \tilde{\nabla}_{a}\tilde{\nabla}^{a}P\nn\\
& +\tfrac{1}{2} \bigl(-4 A_6 + 2 A_8+ A_{10} ( D-1) + 16 b_2 - 32 b_3 - 5 b_4 + 8 b_5 + 16 b_6 - 8 b_7 - 3 b_8 \bigr) \tilde{\nabla}_{a}P \tilde{\nabla}^{a}P \nn\\
&+ \tfrac{1}{8} \bigl(-4 A_6 - 2 A_8+ A_{10} ( D-3) + 32 b_1 - 32 b_3 - 3 b_4 + 8 b_6 - 4 b_7 -  b_8 \bigr) \tilde{\nabla}_{a}\sigma \tilde{\nabla}^{a}\sigma \nonumber\\
\delta P^{1}_A=&\tfrac{1}{8} (-4 A_6 + 2 A_8 + 8 b_2 +32(D-2) b_3 +(D- 5) b_4 -4(D-2) b_5 - 4(3D-7) b_6 \nn\\
&\qquad+4(D- 3) b_7 +(D- 3) b_8 )\tilde{\nabla}_{a}\tilde{\nabla}^{a}\sigma+A_6 \tilde{\nabla}_{a}\sigma \tilde{\nabla}^{a}P  \nonumber\\
\delta g_{Aab}^{1}=&
(8 b_2 - 2 b_4) (\tfrac{1}{2} \tilde{\nabla}_{a}\sigma \tilde{\nabla}_{b}P+ \tfrac{1}{2} \tilde{\nabla}_{a}P\tilde{\nabla}_{b}\sigma)   
+A_8 \tilde{\nabla}_{b}\tilde{\nabla}_{a}\sigma \nonumber \\
&+g_{ab}\Big(\tfrac{1}{2} (- A_{10} + 32 b_3 + b_4 - 4 b_5 - 12 b_6 + 4 b_7 + b_8)\tilde{\nabla}_{c}\tilde{\nabla}^{c}\sigma  
 +A_{10}  \tilde{\nabla}_{c}\sigma \tilde{\nabla}^{c}P \Big)
\labell{Trans1}
\end{align}
The terms with coefficients $A_6,A_8,A_{10}$ are the transformations at order $\alpha'$ which leave the leading $d$-dimensional effective action \reef{LS} to be invariant. As we have shown in the   section 2, the terms with coefficient $A_8$ correspond to the $d$-dimensional coordinate transformations. They have no    effect on our covariant calculations. So we can set $A_8=0$. The other two   coefficients which are not correspond to the $d$-dimensional coordinate transformations, may be fixed at the higher order of $\alpha'$. Our calculations at order $\alpha'^2$, however,  do not fix these coefficients either.  There is only one such unfixed coefficients in the $D$-dimensional field redefinitions at order $\alpha'$.  

Therefore, the T-duality constraint on the effective action fixes the effective action up to  $D$-dimensional field redefinitions and 
up to the overall factor of $b_1$.   The S-matrix calculations fix $b_1=1/4$ for the bosonic theory, $b_1=1/8$ for heterotic theory, and $b_1=0$ for the superstring theory \cite{Metsaev:1986yb}. 


For the particular choice of $b_2=-4b_1,b_3=b_1, b_4=-16b_1,b_5=8b_1,b_6=0,b_7=0,b_8=16b_1$, the  T-duality invariant action \reef{finalS10} becomes
 \beqa
 S_1&=&\frac{-2b_1}{\kappa}\alpha'\int d^{d+1}x e^{-2\Phi}\sqrt{-G}\Big(  R_{\alpha \beta \gamma \delta} R^{\alpha \beta \gamma \delta}-4 R_{\alpha \beta} R^{\alpha \beta} +  R^2  -16 R_{\alpha \beta} \nabla^{\alpha}\Phi \nabla^{\beta}\Phi\nonumber\\
&& + 8 R \nabla_{\alpha}\Phi \nabla^{\alpha}\Phi    + 16 \nabla_{\alpha}\Phi \nabla^{\alpha}\Phi \nabla_{\beta}\nabla^{\beta}\Phi  - 16 \nabla_{\alpha}\Phi \nabla^{\alpha}\Phi \nabla_{\beta}\Phi \nabla^{\beta}\Phi\Big)\labell{SSS}
\eeqa
and the corresponding T-duality transformations \reef{Trans1} become
\begin{align}
\delta\sigma^{1}_A=&\tfrac{1}{8} ( 4 A_6  - A_{10}(D-3) ) \tilde{R} + \tfrac{1}{2} \bigl(4 A_6   -  A_{10} ( D-2) \bigr) \tilde{\nabla}_{a}\tilde{\nabla}^{a}P  \nn\\
&+\tfrac{1}{2} \bigl(-4 A_6  + A_{10} ( D-1)  \bigr) \tilde{\nabla}_{a}P \tilde{\nabla}^{a}P+ \tfrac{1}{8} \bigl(-4 A_6  + A_{10} ( D-3) + 32 b_1   \bigr) \tilde{\nabla}_{a}\sigma \tilde{\nabla}^{a}\sigma \nonumber\\
\delta P^{1}_A=& A_6\Big(-\tfrac{1}{2} \tilde{\nabla}_{a}\tilde{\nabla}^{a}\sigma+  \tilde{\nabla}_{a}\sigma \tilde{\nabla}^{a}P \Big) \nonumber\\
\delta g_{Aab}^{1}=&g_{ab}A_{10}\Big(-\tfrac{1}{2}\tilde{\nabla}_{c}\tilde{\nabla}^{c}\sigma  
 + \tilde{\nabla}_{c}\sigma \tilde{\nabla}^{c}P \Big)\, .
\labell{Trans10}
\end{align}
The effective action \reef{SSS} is the one considered in \cite{Kaloper:1997ux}, and the T-duality transformations \reef{Trans10} for the particular case of $A_6=A_{10}=0$, are those have been found in \cite{Kaloper:1997ux}. 

For the particular choice of $b_4=4b_2$, $b_5=-4b_2-8b_3$, $b_6=2b_2+8b_3$, $b_7=4b_2+16b_3$ and $b_8=-12 b_2-32 b_3$,  the T-duality transformations become \reef{Trans1} and  the effective action,  after using some integrations by part, becomes
\begin{align}
S_1 
=&\frac{-2}{\kappa^2}\alpha'\int d^{D}x e^{-2\Phi}\sqrt{-G}\Big(b_1 R_{\alpha \beta \gamma \delta} R^{\alpha \beta \gamma \delta}+
 b_2\mathcal{R}_{\alpha\beta}^2 +b_3\mathcal{R}^2)\labell{eq.2.135.20}
\end{align}
where $\mathcal{R}_{\alpha\beta}$ and $\mathcal{R}$ are
\beqa
 \mathcal{R}_{\alpha\beta}=R_{\alpha\beta}+2\nabla_\alpha\nabla_\beta\Phi &;& \mathcal{R}=R+4\nabla_\alpha\nabla^\alpha\Phi-4\nabla_\alpha\Phi\nabla^\alpha\Phi\labell{RR}
\eeqa
 It has been shown in \cite{Garousi:2015qgr} that  in the superstring theory, the T-duality invariant effective action of  O$_p$-plane  at order $\alpha'^2$ can be written in terms of $\mathcal{R'}_{\alpha\beta}=R_{\alpha\beta}+\nabla_\alpha\nabla_\beta\Phi$. The reason for the extra factor of 2 for the dilaton in $\mathcal{R}_{\alpha\beta}$ with respect to $\mathcal{R'}_{\alpha\beta}$, is that the overall dilaton factor in the bulk action is $e^{-2\Phi}$ whereas in the brane action is $e^{-\Phi}$. Similarly, in the bosonic string theory, the T-duality invariant effective action of  O$_p$-plane at order $\alpha'$ which has been found in \cite{Garousi:2013gea} can be written in terms of  $\mathcal{R'}=R+2\nabla_a\nabla^a\Phi-\nabla_a\Phi\nabla^a\Phi$, after using an integration by part. 

If one is not interested in the explicit form of the T-duality corrections at order $\alpha'$, the T-duality invariant action at order $\alpha'$ can more easily be found by replacing the $d$-dimensional equations of motion of \reef{LS}, 
into \reef{delS}. Using these equations one can write $\tilde{\nabla}_{a}\tilde{\nabla}^{a}\sigma$, $\tilde{\nabla}_{a}P \tilde{\nabla}^{a}P$ and  $\tilde{R}^{ab}$ in \reef{delS} in terms of other terms, \ie
\begin{align}
&  \tilde{\nabla}_{a}\tilde{\nabla}^{a}\sigma=2\tilde{\nabla}_{a}\sigma \tilde{\nabla}^{a}P \nn\\
& \tilde{\nabla}_{a}P \tilde{\nabla}^{a}P=\frac{1}{4}\tilde{R}+ \tilde{\nabla}_{a}\tilde{\nabla}^{a}P  -\frac{1}{4}\tilde{\nabla}_{a}\sigma \tilde{\nabla}^{a}\sigma\nn\\
& \tilde{R}^{ab} =\tilde{\nabla}^{a}\sigma \tilde{\nabla}^{b}\sigma-2 \tilde{\nabla}^{b}\tilde{\nabla}^{a}P
 \end{align}
Using the total derivative terms \reef{tot} to write independent terms, one can simplify   $\delta\bS_1$ in  \reef{delS} as   
\begin{align}
\delta\bS_1= & -\frac{2}{\kappa^2}\alpha' \int  d^dx e^{-2P}\sqrt{-g} \Big( \tilde{\nabla}_{a}\sigma\tilde{\nabla}^{a}\sigma\tilde{\nabla}_{b}\sigma\tilde{\nabla}^{b}P + 2 \tilde{\nabla}_{a}\tilde{\nabla}^{a}P \tilde{\nabla}_{b}\sigma\tilde{\nabla}^{b}P\Big)\xi 
\end{align}
The T-duality constraint \reef{Teff0} then requires    $\delta\bS_1=0$ which reproduces  the last equation in \reef{c1}.

Using the equations of motion, one may rewrite the above $\delta\bS_1$ as 
\begin{align}
\delta\bS_1= & -\frac{2}{\kappa^2}\alpha' \int  d^dx e^{-2P}\sqrt{-g} \Big( \tilde{\nabla}_{a}\sigma\tilde{\nabla}^{a}\sigma\tilde{\nabla}_{b}\sigma\tilde{\nabla}^{b}P + 4 \tilde{\nabla}_{a}P\tilde{\nabla}^{a}P \tilde{\nabla}_{b}\sigma\tilde{\nabla}^{b}P\Big)\xi 
\end{align}
If one   replaces the T-duality corrections \reef{c1} to the constraint \reef{con1}, one finds exactly the above constraint that has been found by using the $d$-dimensional equations of motion.
   
\subsection{Effective action at order $\alpha'^2$}
 
 In this subsection, we continue the   calculations to find the effective action   at order $\alpha'^2$.
The covariant effective action at order $\alpha'^2$ is constructed from the combinations of the following terms:
\begin{align}
&R_{\alpha\beta\gamma\delta}\ ,
\quad \nabla_\mu R_{\alpha\beta\gamma\delta}\ ,
\quad \nabla_\nu\nabla_\mu R_{\alpha\beta\gamma\delta}\ ,
\quad \nabla_\sigma\nabla_\nu\nabla_\mu R_{\alpha\beta\gamma\delta},
\quad\nabla_\epsilon \nabla_\sigma\nabla_\nu\nabla_\mu R_{\alpha\beta\gamma\delta},\nn\\
& \nabla_\alpha\Phi\ ,
\quad \nabla_\beta\nabla_\alpha\Phi\ ,
\quad\nabla_\gamma\nabla_\beta\nabla_\alpha\Phi\ ,
\quad\nabla_\delta\nabla_\gamma\nabla_\beta\nabla_\alpha\Phi\ ,\nn\\
&\nabla_\mu\nabla_\delta\nabla_\gamma\nabla_\beta\nabla_\alpha\Phi\ ,
\quad\nabla_\nu\nabla_\mu\nabla_\delta\nabla_\gamma\nabla_\beta\nabla_\alpha\Phi\labell{eq.3.10}
\end{align}
Each term must have six derivatives. There are  203 such couplings!
Some of them are related by the  Bianchi identities.
For example:
\begin{align}
\nabla_\alpha R^{\alpha\beta}{}_{\gamma\delta}= \nabla_{\gamma}R^{\beta}{}_{\delta} -  \nabla_{\delta}R^{\beta}{}_{\gamma}
\end{align}
So   each independent term should not contain $\nabla_\alpha R^{\alpha\beta}{}_{\gamma\delta}$. Using these identities, one finds there are only 100 independent terms in the Lagrangian. In the action, however, we are free to drop   total derivative terms. So many of these independent terms are related by total derivative terms.  
For example, up to a total derivative term, we have the following relation:
\begin{align}
&e^{-2 \Phi}\sqrt{-G}\nabla_{\epsilon}R_{\alpha \beta \gamma \delta} \nabla^{\epsilon}R^{\alpha \beta \gamma \delta}= 
-e^{-2\Phi}\sqrt{-G}\Big(2R^{\alpha \beta}R_{\alpha}{}^{\gamma \delta \epsilon}R_{\beta \gamma \delta \epsilon} \nn\\
&- 4R_{\alpha}{}^{\epsilon}{}_{\gamma}{}^{\zeta}R^{\alpha \beta \gamma \delta}R_{\beta \epsilon \delta \zeta} - R_{\alpha \beta}{}^{\epsilon \zeta}R^{\alpha \beta \gamma \delta}R_{\gamma \delta \epsilon \zeta}\nn\\
& - 2R^{\beta \gamma \delta \epsilon} \nabla_{\alpha}R_{\beta \gamma \delta \epsilon} \nabla^{\alpha}\Phi + 4R_{\alpha \gamma \beta \delta} \nabla^{\delta}\nabla^{\gamma}R^{\alpha \beta}\Big)
\end{align}
Using all such relations and the Bianchi identities, one finds there are only 44 independent couplings in the effective action, \ie  
\begin{align}
S_2=&\frac{-2\alpha'^2}{\kappa^2}\int d^{d+1}x\, e^{-2\Phi}\sqrt{-G}\Big(
c_1 R_{\alpha \beta}{}^{\epsilon \zeta} R^{\alpha \beta \gamma \delta} R_{\gamma \delta \epsilon \zeta} + c_2 R_{\alpha}{}^{\epsilon}{}_{\gamma}{}^{\zeta} R^{\alpha \beta \gamma \delta} R_{\beta \epsilon \delta \zeta}  \nonumber \\
& + c_3 R_{\alpha \beta} R^{\alpha \beta} R + c_4 R^3 +c_5 R_{\alpha}{}^{\gamma} R^{\alpha \beta} R_{\beta \gamma}+ c_6 \nabla_{\gamma}R_{\alpha \beta} \nabla^{\gamma}R^{\alpha \beta} + c_7 R R_{\alpha \beta \gamma \delta} R^{\alpha \beta \gamma \delta} \nonumber \\ 
& + c_8 R^{\alpha \beta} R_{\alpha}{}^{\gamma \delta \zeta} R_{\beta \gamma \delta \zeta} + c_9 R^{\alpha \beta} R^{\gamma \delta} R_{\alpha \gamma \beta \delta} + c_{10} \nabla_{\alpha}R \nabla^{\alpha}R+ c_{11} \nabla_{\alpha}\nabla^{\alpha}\Phi \nabla_{\beta}\nabla^{\beta}\Phi \nabla_{\gamma}\nabla^{\gamma}\Phi \nonumber \\ 
&+ c_{12} \nabla_{\alpha}\Phi \nabla^{\alpha}\Phi \nabla_{\beta}\nabla^{\beta}\Phi \nabla_{\gamma}\nabla^{\gamma}\Phi + c_{13} \nabla_{\alpha}\Phi \nabla^{\alpha}\Phi \nabla_{\beta}\Phi \nabla^{\beta}\Phi \nabla_{\gamma}\nabla^{\gamma}\Phi  \nonumber \\ 
& + c_{14} \nabla_{\alpha}\Phi \nabla^{\alpha}\Phi \nabla_{\beta}\Phi \nabla^{\beta}\Phi \nabla_{\gamma}\Phi \nabla^{\gamma}\Phi + c_{15} \nabla_{\alpha}\nabla^{\alpha}\Phi \nabla_{\gamma}\nabla^{\gamma}\nabla_{\beta}\nabla^{\beta}\Phi\nonumber \\ 
&  + c_{16} \nabla_{\alpha}\Phi \nabla^{\alpha}\Phi \nabla_{\gamma}\nabla^{\gamma}\nabla_{\beta}\nabla^{\beta}\Phi + c_{17} \nabla_{\alpha}\nabla^{\alpha}\Phi \nabla_{\gamma}\nabla_{\beta}\Phi \nabla^{\gamma}\nabla^{\beta}\Phi + c_{18} \nabla_{\alpha}\Phi \nabla^{\alpha}\Phi \nabla_{\gamma}\nabla_{\beta}\Phi \nabla^{\gamma}\nabla^{\beta}\Phi \nonumber \\ 
& + c_{19} R \nabla_{\alpha}\Phi \nabla^{\alpha}\Phi \nabla_{\beta}\Phi \nabla^{\beta}\Phi + c_{20} R \nabla_{\alpha}\Phi \nabla^{\alpha}\Phi \nabla_{\beta}\nabla^{\beta}\Phi + c_{21} R \nabla_{\alpha}\nabla^{\alpha}\Phi \nabla_{\beta}\nabla^{\beta}\Phi  \nonumber\nonumber \\ 
&+ c_{22} R \nabla_{\beta}\nabla^{\beta}\nabla_{\alpha}\nabla^{\alpha}\Phi + c_{23} R \nabla_{\beta}\nabla_{\alpha}\Phi \nabla^{\beta}\nabla^{\alpha}\Phi  + c_{24} R \nabla^{\alpha}\Phi \nabla_{\beta}\nabla_{\alpha}\Phi \nabla^{\beta}\Phi  \nonumber \\ 
&+ c_{25} R \nabla^{\alpha}\Phi \nabla_{\beta}\nabla^{\beta}\nabla_{\alpha}\Phi  + c_{26} R_{\alpha \beta} \nabla^{\alpha}\Phi \nabla^{\beta}\Phi \nabla_{\gamma}\nabla^{\gamma}\Phi+ c_{27} R_{\beta \gamma} \nabla_{\alpha}\Phi \nabla^{\alpha}\Phi \nabla^{\beta}\Phi \nabla^{\gamma}\Phi\nonumber\\
&+ c_{28} R^{\alpha \beta} \nabla_{\beta}\nabla_{\alpha}\Phi \nabla_{\gamma}\nabla^{\gamma}\Phi + c_{29} R^{\beta \gamma} \nabla_{\alpha}\Phi \nabla^{\alpha}\Phi \nabla_{\gamma}\nabla_{\beta}\Phi + c_{30} R^{\alpha \beta} \nabla_{\gamma}\nabla_{\beta}\Phi \nabla^{\gamma}\nabla_{\alpha}\Phi \nonumber\\
& + c_{31} R_{\alpha \gamma \beta \delta} \nabla^{\alpha}\Phi \nabla^{\beta}\Phi \nabla^{\delta}\nabla^{\gamma}\Phi+ c_{32} R^2 \nabla_{\alpha}\Phi \nabla^{\alpha}\Phi + c_{33} R^2 \nabla_{\alpha}\nabla^{\alpha}\Phi + c_{34} R_{\beta \gamma} R^{\beta \gamma} \nabla_{\alpha}\Phi \nabla^{\alpha}\Phi \nonumber\\
&+ c_{35} R_{\alpha \beta} R^{\alpha \beta} \nabla_{\gamma}\nabla^{\gamma}\Phi + c_{36} R_{\alpha \beta} R \nabla^{\alpha}\Phi \nabla^{\beta}\Phi + c_{37} R^{\alpha \beta} R \nabla_{\beta}\nabla_{\alpha}\Phi + c_{38} R_{\alpha}{}^{\gamma} R_{\beta \gamma} \nabla^{\alpha}\Phi \nabla^{\beta}\Phi \nonumber \\ 
& + c_{39} R_{\alpha}{}^{\gamma} R^{\alpha \beta} \nabla_{\gamma}\nabla_{\beta}\Phi  + c_{40} R^{\gamma \delta} R_{\alpha \gamma \beta \delta} \nabla^{\alpha}\Phi \nabla^{\beta}\Phi+ c_{41} R^{\alpha \beta} R_{\alpha \gamma \beta \delta} \nabla^{\delta}\nabla^{\gamma}\Phi \nonumber \\ 
& + c_{42} R_{\beta \gamma \delta \zeta} R^{\beta \gamma \delta \zeta} \nabla_{\alpha}\Phi \nabla^{\alpha}\Phi + c_{43} R_{\beta \gamma \delta \zeta} R^{\beta \gamma \delta \zeta} \nabla_{\alpha}\nabla^{\alpha}\Phi  + c_{44} R_{\alpha}{}^{\gamma \delta \zeta} R_{\beta \gamma \delta \zeta} \nabla^{\alpha}\Phi \nabla^{\beta}\Phi 
\Big)\labell{eq.3.20}
\end{align}
where $c_1,\cdots, c_{44}$ are the unknown coefficients. 
Apart from the coefficients $c_1,c_2$ which are invariant under field redefinitions, all other coefficients are a priori ambiguous because they are changed under the field redefinitions. It has been shown in \cite{Bento1989,Bento1990} that there are five different combinations of the ambiguous coefficients that are invariant under the field redefinitions. Considering the  transformation of effective action $ S_0+S_1 $ under the general field redefinitions $ G_{\mu\nu}\to G_{\mu\nu}+ \alpha' \delta G^{(1)}_{\mu\nu}+\alpha'^2\delta G^{(2)}_{\mu\nu}+\cdots$ and $ \Phi\to \Phi+\alpha'\delta\Phi^{(1)}+\alpha'^2\delta\Phi^{(2)}+\cdots $, \ie
\beqa
S_0+S_1&\rightarrow &S_0+S_1+\alpha'\frac{\delta S_1}{\delta G } \delta G^{(1)} +\alpha'\frac{\delta S_1}{\delta \Phi} \delta \Phi^{(1)}\nn\\
&&
 +\alpha'^2\frac{\delta S_0}{\delta G } \delta G^{(2)} +\alpha'^2\frac{\delta S_0}{\delta \Phi} \delta \Phi^{(2)}+\alpha'^2S_0(\delta G^{(1)},\delta\Phi^{(1)})+\cdots\label{FRS2}
\eeqa
where $ S_0 $ and $ S_1 $ are  the effective actions  at order $ \alpha'^0 $ and $ \alpha' $, respectively. When one replaces the field redefinitions in the effective action $S_0$, one finds  terms which have $\delta G^{(1)}\delta G^{(1)}$, $\delta G^{(1)}\delta \Phi^{(1)}$ and $\delta \Phi^{(1)}\delta \Phi^{(1)}$. The expression $S_0(\delta G^{(1)},\delta\Phi^{(1)})$ represents these terms. These terms cause  the field redefinitions at order $\alpha'^2$ not to be identical to all possible substitutions of lower order equations of motion, \ie $\frac{\delta S_0}{\delta G }=\frac{\delta S_1}{\delta G } =\frac{\delta S_0}{\delta \Phi }=\frac{\delta S_1}{\delta \Phi } =0$, in the effective action at order $\alpha'^2$.

Considering  all possible terms for $\delta G$ and $\delta\Phi$, one finds the following combinations of the coefficients remain invariant under the field redefinitions\footnote{  There is an extra factor of $8\delta c_6$ in the fifth equation in \cite{Bento1989} that our calculation does not produce it. We think it should be a typo in \cite{Bento1989}.}:
\beqa
  \xi_1&\equiv&b_1(24   b_3 + \tfrac{1}{2}   b_4 - 3   b_5 - 8   b_6 + 2  b_7 + \tfrac{1}{2}  b_8)
-4   c_7 +   c_{42} + 2   c_{43},\nonumber \\
   \xi_2&\equiv& b_1 (8  b_2 - 2    b_4)+2  c_4 - 4   c_6 -   c_{41} +   c_{44},\nonumber\\
  \xi_3&\equiv&-8   c_5 - 2   c_{30} -   c_{31} + 4   c_{39} + 2   c_{40}, \nonumber\\
  \xi_4&\equiv&-\frac{1}{2} (16   b_2 + 48   b_3 - 3   b_4 - 6   b_5 - 16   b_6 + 4  b_7 +   b_8)\xi\nonumber\\
&&+ 64   c_9 - 8   c_{11} - 4   c_{12} - 2  c_{13} -   c_{14} + 4   c_{19} + 8   c_{20} + 16  c_{21} - 
   16   c_{32} - 32  c_{33} ,\nonumber\\
  \xi_5&\equiv& 4 b_2^2 + 64 b_3^2 + \tfrac{5}{8} b_4^2 -  b_4 b_5 + b_5^2 - 4 b_4 b_6 + 6 b_5 b_6 + 10 b_6^2 + 2 b_4 b_7 - 2 b_5 b_7 - 8 b_6 b_7 + 2 b_7^2  \nonumber\\
&&+ \tfrac{1}{2} b_4 b_8-  \tfrac{1}{2} b_5 b_8 - 2 b_6 b_8 + b_7 b_8 + \tfrac{1}{8} b_8^2 + 4 b_3 (2 b_4 - 4 b_5 - 12 b_6 + 4 b_7 + b_8)\nonumber\\
&&  -  b_2 (16 b_3 + 3 b_4 - 2 b_5 - 8 b_6 + 4 b_7 + b_8)-  \tfrac{1}{16} (32 b_3 + b_4 - 4 b_5 - 12 b_6 + 4 b_7 + b_8)^2 D\nn\\
&&   - 8 c_8+ 16 c_{10} + 4 c_{15} + 2 c_{16} + c_{17} + \tfrac{1}{2} c_{18} - 8 c_{22} - 2 c_{23} - 2 c_{28}-  c_{29} + 2 c_{34} + 4 c_{35}   \nn\\
&&+ 4 c_{37}-c_{38}  \labell{5rel}
\eeqa
Hence, the field redefinition freedom allows one to set 37 ambiguous coefficients zero. The S-matrix calculation, then fixes $c_1,c_2$ and the other 5 coefficients that are invariant under the field redefinition \cite{Bento1989,Bento1990}. The values of these   constants depend on which effective action is used at order $\alpha'$. When one uses the effective action \reef{SSS}, the S-matrix fixes $c_1=-\frac{3}{4}c_2\neq 0$ and all other $c$-coefficients to be zero \cite{Bento1990}. Replacing the corresponding $b$-coefficients, \ie $b_2=-4b_1,b_3=b_1, b_4=-16b_1,b_5=8b_1,b_6=0,b_7=0,b_8=16b_1$,  in \reef{5rel}, one finds $\xi_1=\xi_2=\xi_3=\xi_4=\xi_5=0$. Since these functions are invariant under the field redefinitions, in any other field variables these functions are also zero.

We are going, however,  to find the effective action  from the T-duality constraint \reef{deltaS1S2}. So we calculate $\delta \bS_2$ which is $\alpha'^2$-terms resulted from the transformation of d-dimensional action  $\bS_0+\bS_1$ where $\bS_0$ is \reef{LS} and $\bS_1$ is reduction of action \reef{finalS10}, under the T-duality transformation $T^{(0)}+\alpha'T^{(1)}+\alpha'^2T^{(2)}$.   They must be equated with $\bS_2-\bS_2'$ where $\bS_2$ is  reduction of the action \reef{eq.3.20} and $\bS_2'$ is its transformation under the Buscher rules. This equality which is extension of \reef{con} to order $\alpha'^2$, produces some constraints on the coefficients in \reef{eq.3.20}. After subtracting some total derivative terms to finds independent constraints, we have found that there are 67 relations.  One of them is 
\beqa
c_2 \to  -\frac{4}{3}c_1\labell{c12}
\eeqa
which is a relation between the  unambiguous coefficients. There are five relations between the ambiguous coefficients,\ie
\begin{align}
\xi_1=\xi_2=\xi_3=\xi_4=\xi_5=0
\labell{alpha22}
\end{align}
The relations \reef {c12} and \reef{alpha22} are exactly the relations that one finds from the S-matrix calculations. There are also 61 relations which relate  61 $A$-coefficients in \reef{T2}  in terms of other 37 $A$-coefficients at order $\alpha'^2$ , $b$-coefficients , $c$-coefficients, the dimension of spacetime and the residual T-duality parameters $A_6, A_{10}$ at order $\alpha'$. 
They are very lengthy expressions, so  we do not write  the form of the T-duality transformations. 
 
 If one uses the relations \reef{alpha22} to write $c_{18}, c_{19},c_{31}, c_{42}, c_{44}$ in terms of all other ambiguous coefficients, and set all the remaining $c$-coefficients to be zero, the T-duality invariant effective action becomes 
\beqa
S_2&=&\frac{-2\alpha'^2}{\kappa^2}\int d^{d+1}x\, e^{-2\Phi}\sqrt{-G}\Bigg[
c_1 R_{\alpha \beta}{}^{\epsilon \zeta} R^{\alpha \beta \gamma \delta} R_{\gamma \delta \epsilon \zeta} -\frac{3}{4}c_1 R_{\alpha}{}^{\epsilon}{}_{\gamma}{}^{\zeta} R^{\alpha \beta \gamma \delta} R_{\beta \epsilon \delta \zeta}  \nonumber \\
&&- \tfrac{1}{2} b_1 (48 b_3 + b_4 - 6 b_5 - 16 b_6 + 4 b_7 + b_8) R_{\beta \gamma \delta \zeta} R^{\beta \gamma \delta \zeta} \nabla_{\alpha}\Phi \nabla^{\alpha}\Phi \nn\\
&&+ 2 b_1 (-4 b_2 + b_4) R_{\alpha}{}^{\gamma \delta \zeta} R_{\beta \gamma \delta \zeta} \nabla^{\alpha}\Phi \nabla^{\beta}\Phi +\tfrac{1}{8} \nabla_{\alpha}\Phi \nabla^{\alpha}\Phi \nabla_{\gamma}\nabla_{\beta}\Phi \nabla^{\gamma}\nabla^{\beta}\Phi\biggl(-64 b_2^2 \nn\\
&&+ 16 b_2 (16 b_3 + 3 b_4 - 2 b_5 - 8 b_6 + 4 b_7 + b_8)+ (32 b_3 + b_4 - 4 b_5 - 12 b_6 + 4 b_7 + b_8)^2 D \nn\\
&&- 2 \Big[512 b_3^2 + 5 b_4^2 + 8 \bigl(b_5^2 + 6 b_5 b_6 + 10 b_6^2 - 2 (b_5 + 4 b_6) b_7 + 2 b_7^2\bigr)- 4 (b_5 + 4 b_6 - 2 b_7) b_8+ b_8^2\nn\\
&&  + 32 b_3 (2 b_4 - 4 b_5 - 12 b_6 + 4 b_7 + b_8) + 4 b_4 (-2 b_5 - 8 b_6 + 4 b_7 + b_8)\Big]\biggr)\labell{S22}
\Bigg]
\eeqa
Note that the coefficients $c_{19},c_{31}$ become zero. 
As can be seen, the form of the T-duality invariant action $S_2$ depends on the form of action at order $\alpha'$. 

To compare the above action with the actions at order $\alpha'^2$ in the literature, we choose the T-duality invariant action $S_1$ to be \cite{Bento:1989ir}
 \beqa
 S_1&=&\frac{-2b_1}{\kappa}\alpha'\int d^{d+1}x e^{-2\Phi}\sqrt{-G}\Big(  R_{\alpha \beta \gamma \delta} R^{\alpha \beta \gamma \delta}-4 R_{\alpha \beta} R^{\alpha \beta} +  R^2  -16D\frac{D-3}{(D-2)^2} R_{\alpha \beta} \nabla^{\alpha}\Phi \nabla^{\beta}\Phi\nonumber\\
&& + 8D\frac{D-3}{(D-2)^2}  R \nabla_{\alpha}\Phi \nabla^{\alpha}\Phi    + 16\frac{(D-3)(D+2)}{(D-2)^2}  \nabla_{\alpha}\Phi \nabla^{\alpha}\Phi \nabla_{\beta}\nabla^{\beta}\Phi\nn\\
&&  - 16\frac{D^2-8}{(D-2)^2}  \nabla_{\alpha}\Phi \nabla^{\alpha}\Phi \nabla_{\beta}\Phi \nabla^{\beta}\Phi\Big)
\eeqa
The corresponding T-duality invariant action at order $\alpha'^2$ is given by \reef{S22} in which 
\beqa
  c_{42}&=&\frac{D-6}{2(D-2)^2}\nn\\
  c_{44}&=&-\frac{2(D-4)}{(D-2)^2}\nn\\
 c_{18}&=&-5\frac{(D-3)(D-6)}{(D-2)^4}
\eeqa
These coefficients are exactly those found in \cite{Bento:1989ir} by S-matrix and $\sigma$-model calculations. 

 If one chooses the action at order $\alpha'$ to be \reef{SSS}, then the action at order $\alpha'^2$ becomes  
\beqa
S_2&=&\frac{-2\alpha'^2c_1}{\kappa^2}\int d^{d+1}x\, e^{-2\Phi}\sqrt{-G}\Big(
 R_{\alpha \beta}{}^{\epsilon \zeta} R^{\alpha \beta \gamma \delta} R_{\gamma \delta \epsilon \zeta} -\frac{4}{3} R_{\alpha}{}^{\epsilon}{}_{\gamma}{}^{\zeta} R^{\alpha \beta \gamma \delta} R_{\beta \epsilon \delta \zeta}  \Big)\labell{final}
\eeqa
 The   action \reef{final} is exactly the action that has been found in \cite{Metsaev:1987zx,Jack:1988rq} from the S-matrix calculation. The S-matrix fixes $c_1=1/16$ in the bosonic theory and $c_1=0$ in the heterotic and the superstring theories. 

In the heterotic theory,   there are also couplings at order $\alpha'^2$ which are resulted from the Green-Schwarz mechanism \cite{Green:1984sg}. In the supergravity at the leading order of $\alpha'$, the B-field 
  strength $ H(B) $ must be replaced by the improved field
strength $ \widehat{H}(B,\Gamma) $ that  includes  the  Chern-Simons term  built  from  the  Christoffel  connection:
\begin{align}
\widehat{H}_{\mu\nu\rho}(B,\Gamma)=3(\partial_{[\mu}B_{\nu\rho]}+\alpha'\Omega(\Gamma)_{\mu\nu\rho})
\end{align}
with the Chern-Simons three-form
\begin{align}
\Omega(\Gamma)_{\mu\nu\rho}=\Gamma^\alpha_{[\mu|\beta|}\partial_\nu\Gamma^\beta_{\rho]\alpha}+\frac{2}{3}\Gamma^\alpha_{[\mu|\beta|}\Gamma^\beta_{\nu|\gamma|}\Gamma^\gamma_{\rho]\alpha}\,.
\end{align}
Reducing   $ \Omega^2$ from 10-dimensional to 9-dimensional spacetime, one would find   no term which has $ \sigma $.
As a result, one finds,  when metric is diagonal and B-field is zero, $\alpha'^2\Omega^2$ is invariant under the Buscher rules.

We have seen that the T-duality  does  not transform the Riemann curvature couplings to the couplings involving the Ricci curvature, scalar curvature or the dilaton, \ie $\xi$'s are zero. We expect this property for the T-duality at all higher orders of $\alpha'$. This may indicate that in the string frame there is a scheme in which there is no  Ricci or the scalar curvatures and the dilaton appears in the effective action only through the overall dilaton factor . In this scheme, one may use  the T-duality invariance of the effective action that we have used in this paper,  to find only the Riemann curvature couplings. It would be interesting to perform this calculation at order $\alpha'^3$ to find the Riemann curvature couplings at order $\alpha'^3$ that are known in the literature.

We have assumed in this paper that the B-field is zero and the metric is diagonal. The   covariance form of the gravity couplings ensures that they are correct couplings for the general metric. It would be interesting to extend the T-duality invariant  effective actions that we have found in this paper to include the B-field. The B-field corrections at order $\alpha'$ to the action \reef{SSS} and its corresponding T-duality transformations have  been found in \cite{Kaloper:1997ux}. The DFT formulation of the effective action at order $\alpha'$ has been also found in \cite{Marques:2015vua,Baron:2017dvb}.  
 
{\bf Acknowledgments}: H.R. would like to thank G. Jafari for helping with ``xAct" package.  This work is supported by Ferdowsi University of Mashhad under grant  3/31999(1393/07/02).

\end{document}